\documentclass{JHEP3} 
\usepackage{epsfig,bm}

\newcommand{\tr}{\mathop{\rm tr}}
\newcommand{\vol}{\mathop{\rm vol}}
\newcommand{\relint}{\mathop{\rm relint}}
\newcommand{\conv}{\mathop{\rm conv}}

\newcommand{\C}{{\mathbb{C}}}
\newcommand{\R}{{\mathbb{R}}}
\newcommand{\Q}{{\mathbb{Q}}}
\newcommand{\Z}{{\mathbb{Z}}}
\newcommand{\T}{{\mathcal{T}}}
\newcommand{\Tg}{{\mathcal{T}_{\mathrm{gen}}}}
\newcommand{\cZ}{{\mathcal{Z}}}
\newcommand{\vev}[1]{\left\langle #1 \right\rangle}
\renewcommand{\v}[1]{\boldsymbol{#1}}

\newtheorem{theorem}{Theorem}[section]
\newtheorem{corollary}[theorem]{Corollary}
\newtheorem{lemma}[theorem]{Lemma}
\newtheorem{proposition}[theorem]{Proposition}
\newcommand{\qed}{$\Box$}

\title{Zonotopes and four-dimensional superconformal field theories}
\author{Akishi Kato\thanks{Partially Supported by Grants-in-Aid for 
Scientific Research and the Japan Society for Promotion of Science (JSPS)}
\\
Graduate School of Mathematical Sciences, University of Tokyo\\
3-8-1 Komaba, Meguro-ku, Tokyo 153-8914, Japan.\\
E-mail: \email{akishi@ms.u-tokyo.ac.jp}}

\abstract{The $a$-maximization technique proposed by Intriligator and
Wecht allows us to determine the exact $R$-charges and scaling
dimensions of the chiral operators of four-dimensional superconformal
field theories.  The problem of existence and uniqueness of the
solution, however, has not been addressed in general setting.  In this
paper, it is shown that the $a$-function always has a unique critical
point which is also a global maximum for a large class of quiver gauge
theories specified by toric diagrams. Our proof is based on the
observation that the $a$-function is given by the volume of a three
dimensional polytope called ``zonotope'', and the uniqueness essentially
follows from Brunn-Minkowski inequality for the volume of convex bodies. 
We also show a universal upper bound for the exact $R$-charges, and the
monotonicity of $a$-function in the sense that $a$-function decreases
whenever the toric diagram shrinks. The relationship between
$a$-maximization and volume-minimization is also discussed.  }


\preprint{hep-th/0610266, UTMS 2006-28}

\begin{document}

\section{Introduction}

One of the most important problems in quantum field theories is to
understand the renormalization group (RG) flows and the universality
classes.

In two dimensions, we have a fairly satisfactory global picture of the
moduli space $\mathcal{M}_{\mathrm{2d QFT}}$ of quantum field
theories. Zamolodchikov \cite{MR865077} introduced a real valued
function $c:\mathcal{M}_{\mathrm{2d QFT}}\to \R$ and showed that the RG
flow is a gradient flow of $c$ with respect to the metric defined by
two-point correlation functions. In particular, $c$ is monotonically
decreasing along the RG flow.  Each critical point of $c$ corresponds to
a fixed point of the RG flow i.e. a conformal field theory, and the
critical value is the central charge of the Virasoro algebra of the
corresponding conformal field theory.

Considerable effort has been expended to generalize these ideas to to
four dimensions. As the Zamolodchikov's $c$-function is related the
trace anomaly of the stress energy tensor, natural candidates for four
dimensional theories are the coefficients $a$ and $c$ of trace anomaly
 \cite{Duff:1993wm}
\[
 g_{ij} \vev{T^{ij}} = -a E_{4} - c I_{4} 
\]
where
\begin{eqnarray*}
  E_{4}&=&\frac{1}{16\pi^{2}}\bigl(R^{ijkl}R_{ijkl}-4R^{ij}R_{ij}+R^{2}
  \bigr),\nonumber\\
  I_{4}&=&-\frac{1}{16\pi^{2}}\bigl(R^{ijkl}R_{ijkl}
  -2R^{ij}R_{ij}+\frac{1}{3}R^{2}
  \bigr),
\end{eqnarray*}
where $R_{ijkl}$ denotes the Riemann curvature of the background
geometry.  It is now believed \cite{MR973592} that $a$-function will
play a similar role to Zamolodchikov's $c$ function: $a$ decreases along
any RG flow.

It is usually difficult to compute $a$-functions. The situation is much
better if the field theories has supersymmetry. Any four dimensional
superconformal fields theory (SCFT) has global symmetry supergroup
$SU(2,2|1)$; $SO(4,2)\times U(1)_{R}$ is its bosonic subgroup. For the
representation of the superconformal algebra on the chiral
supermultiplet \cite{MR812225,MR740818}, there is a simple relation
between the $R$-charge and the conformal dimension $\Delta$ of a
operator $\mathcal{O}$
\[
  \Delta(\mathcal{O}) = \frac{3}{2} R(\mathcal{O}).
\]
The scaling dimensions of chiral operators are protected from quantum
corrections.  Anselmi et al.  \cite{Anselmi:1997ys,Anselmi:1997am} have
shown that the $U(1)_{R}$ {}'t Hooft anomalies completely determine the
$a$ and $c$ central charges of the superconformal field theory:
\[
  a=\frac{3}{32}\bigl(3\tr R^{3}-\tr R\bigr),
  \qquad 
 c=\frac{1}{32}\bigl(9\tr R^{3}-5\tr R\bigr).
\]
Here $R$ denotes the generator of the $U(1)_{R}$ symmetry and the traces
are taken over all the fields in the field theory. Thus $U(1)_{R}$
symmetry is extremely useful if correctly identified; it is in general,
however, a nontrivial linear combination of all non-anomalous global
$U(1)$ symmetries.

The crucial observation by Intriligator and Wecht
\cite{Intriligator:2003jj} is that the correct combination should be
free of Adler-Bell-Jackiw type anomalies i.e. the NSVZ exact beta
functions \cite{Novikov:1983uc} vanish for all gauge groups. Denoting by
$F_{1},\dots,F_{n}$ the global charges of non-anomalous $U(1)$
symmetries, the conditions are
\begin{equation}
 \label{eq:VanishingBeta}
 \left\{
  \begin{array}{ll}
   & 9 \tr R^{2}F_{i} = \tr F_{i} \qquad (i=1,\dots,n) \\
   & (\tr R F_{j} F_{k} )_{j,k=1}^{n} \;: \;\mbox{negative definite}
  \end{array}
 \right.
\end{equation}
where the second line is required by the unitarity of the conformal
field theory. These conditions are succinctly stated as ``exact $U(1)_R$
charges maximize $a$'':
\begin{theorem}[\cite{Intriligator:2003jj}]
 Among all possible combination of abelian currents
 \[
     R_{\phi}=R_{0}+\sum_{i=1}^{n}\phi^{i} F_{i},
 \]
 the correct $U(1)_{R}$ current is given by the $\phi$ which attains a
 local maximum of the ``trial'' $a$-function
 \[
    a(\phi)=\frac{3}{32}\bigl(3\tr R_{\phi}^{3}-\tr R_{\phi}\bigr).
 \]
\end{theorem}

It is thus quite natural and important to investigate the existence and
uniqueness of the solution to the $a$-maximization. By continuity of the
trial $a$-function, a maximizer always exists on every closed set. But
this may not be a critical point; the anomaly-free condition
(\ref{eq:VanishingBeta}) requires that the point should be critical.  On
the other hand, if there are several local maxima, which one gives the
``correct'' $R$-charges? If there is a critical point which is not local
maximum i.e. saddle point, what happens? How does the change of toric
diagrams influence the maxima of trial $a$-functions? To the best
of the authors knowledge, however, no general answer to these questions
is known.

The purpose of this paper is to answer these questions. We prove that
the $a$-function has always a unique critical point which is also a
global maximum for a large class of quiver gauge theories specified by
toric diagrams, i.e. two dimensional convex polygons. The monotonicity
of $a$-function is also established in the sense that $a$-function
decreases whenever the toric diagram shrinks. We derive these results
purely mathematically, although the setting of the problem is
substantially based on the conjectural AdS/CFT correspondence or
gauge/gravity duality. Hopefully, our result will be useful toward the
proof of these conjectures.

The organization of the paper is as follows. In section 2, we briefly
summarize the rule how a toric diagram determines the trial $a$-function
of a quiver gauge theory.  Section 3 is devoted to set up a mathematical
framework of $a$-maximization and state our main theorems. In section 4,
we observe that the $a$-function is given by the volume of a three
dimensional polytope called ``zonotope''; the uniqueness of the critical
point then follows from Brunn-Minkowski inequality as we discuss in
section 5. In Section 6 we show the existence of the critical point,
i.e. the solution to the $a$-maximization. In Section 7, we derive a
universal upper bound on $R$-charges using the interpretation as a
volume. The monotonicity of $a$-function is established in Section 8.
In section 9, the relationship between $a$-maximization and volume
minimization proposed by Martelli, Sparks and Yau
\cite{Martelli:2005tp,Martelli:2006yb} is discussed. In particular, the
Reeb vector is shown to be pointing to the zonotope center and the
results of Butti and Zaffaroni \cite{Butti:2005vn} is rederived. In the
final section the results are summarized and a short outlook is given.

\section{Toric diagrams and $a$-functions}
\label{sec:A-from-P}

There is a general formula for $a$-functions based only on toric
diagrams, which we summarize below. For more details we refer the reader
to \cite{Hanany:2001py,Benvenuti:2004dy,Butti:2005vn,Franco:2005rj,%
Benvenuti:2005ja} and references therein.

\FIGURE{\epsfig{file=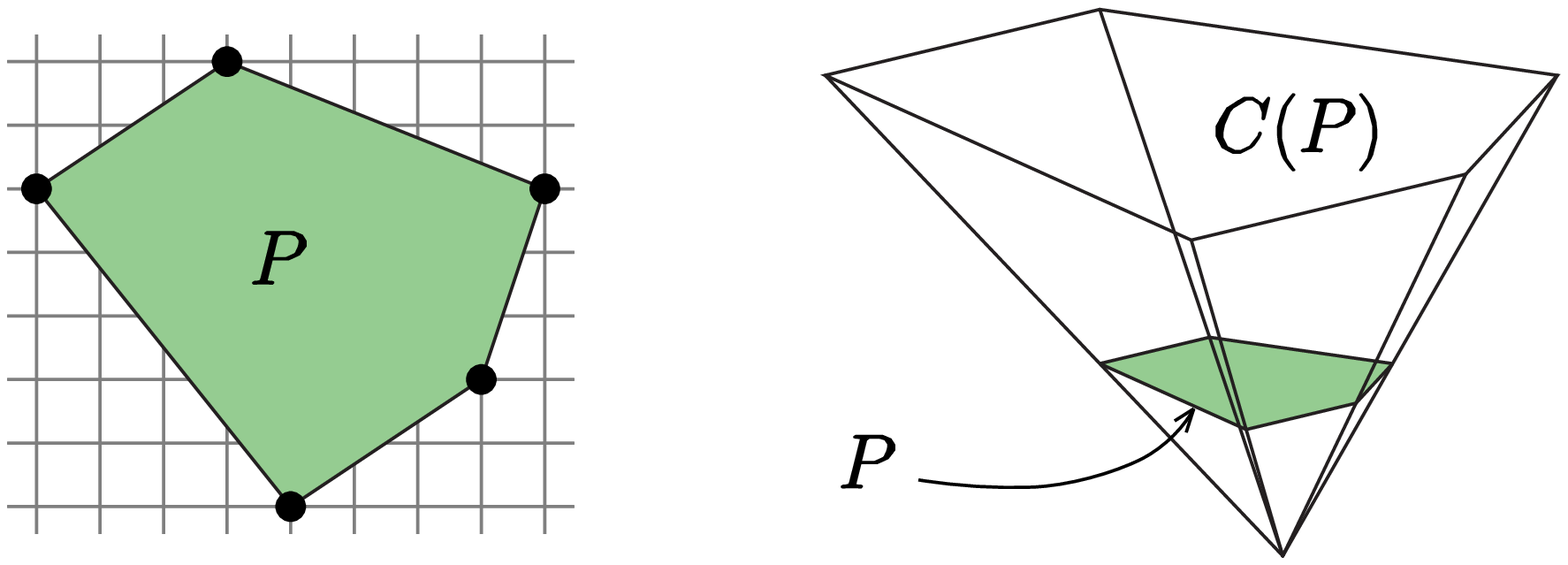,width=.60\textwidth}
 \caption{Toric diagram $P$ and the cone $C(P)$
 \label{fig:toric-diag-cone}}
}

A \emph{toric diagram} $P\subset \mathbb{R}^{3}$ is a two dimensional
convex integral polygon embedded into height one; the coordinates of
each vertex is of the form $(*,*,1)\in \mathbb{Z}^{3}$ (Figure
\ref{fig:toric-diag-cone}). Let $\T$ denote the set of toric diagrams.
For an $n$-gon $P\in \T$, denote its vertices\footnote{As usual, we assume
that all the vertices are extremal points of $P$. A point $\v{v}\in P$
is called an \emph{extremal point} of a polytope $P$ if $\v{v}$ cannot
be expressed as $\alpha \v{a}+\beta\v{b}$ where $\v{a},\v{b}$ are
distinct points in $P$ and $\alpha,\beta$ are positive numbers such that
$\alpha+\beta=1$. }  by $\v{v}_{1},\v{v}_{2},\dots,\v{v}_{n}$ in
counter-clockwise order, so that
\[
  \vev{\v{v}_{i},\v{v}_{j},\v{v}_{k}} > 0, \qquad (1\leq i<j<k\leq n). 
\]
Here and throughout this paper, $\vev{\v{u},\v{v},\v{w}}$ denotes the
determinant of the $3\times 3$ matrix whose columns are
$\v{u},\v{v},\v{w}\in \R^{3}$.  We adopt the convention that the indices
are defined modulo $n$, $\v{v}_{i}=\v{v}_{i+n}$; thus
$\vev{\v{v}_{i-1},\v{v}_{i},\v{v}_{i+1}}>0$ for all $i$. The cone $C(P)$
over the base $P$ defined by
\[
  C(P):= \R_{\geq 0}\v{v}_{1}+\dots+\R_{\geq 0}\v{v}_{n}.
\]
will be important for the relation with volume minimization (see Section
\ref{sec:vol-min}).

With each toric diagram $P\in \T$ there is associated a quiver gauge
theory.  A quiver is a directed graph encoding a gauge theory which
gives rise to a SCFT. For our purpose, however, a specific form of the
quiver is not needed. Let $N_\mathrm{gauge}$ and $N_\mathrm{matter}$
denote the number of the vertices and the edges of the quiver,
respectively. Each vertex is in one-to-one correspondence with a $U(N)$
factor of the total gauge group $U(N)^{N_\mathrm{gauge}}$; each edge
represents a chiral bi-fundamental field.  The numbers
$N_\mathrm{gauge}$ and $N_\mathrm{matter}$ can be extracted directly
from the toric diagram $P$:
\begin{eqnarray}
   N_\mathrm{gauge}&
   =&\sum_{1\leq i\leq n} \vev{\v{v}_{i},\v{v}_{i+1},\v{e}_{3}}
   =2\mathrm{Area}(P),
   \nonumber
   \\
   N_\mathrm{matter}& =&\sum_{1\leq i<j\leq n} 
   |\vev{\v{v}_{i}-\v{v}_{i-1},\v{v}_{j}-\v{v}_{j-1},\v{e}_{3}}|.
 \label{eq:Ns-from-P}
\end{eqnarray}
Here $|~~|$ denotes the usual absolute value and $\v{e}_{3}$ is the unit
vector $(0,0,1)\in \Z^{3}$.

\FIGURE[b]{
 \begin{minipage}[c]{.49\textwidth}
  \centering
  \epsfig{file=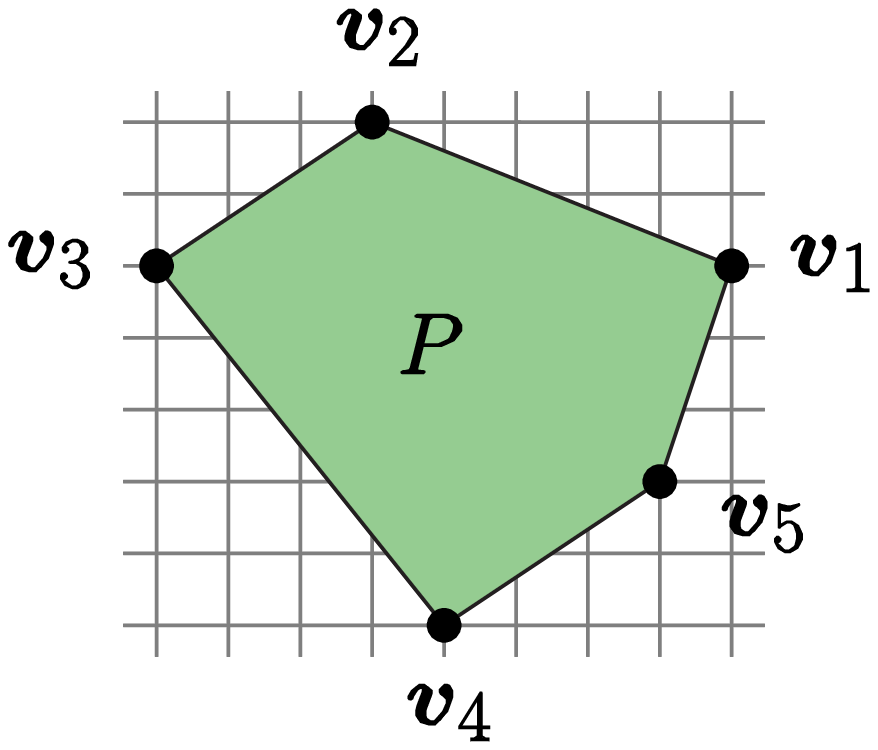,width=.6\textwidth}
 \end{minipage}
 \begin{minipage}[c]{.49\textwidth}
  \begin{small}
  \[
    \begin{array}{|c|c|c|}
  \hline
  (i,j)& \mbox{multiplicity}& \mbox{R-charge}\\
  \hline
  (1,2)& 17& \phi^{1}\\
  \hline
  (1,3)& \phantom{0}7& \phi^{1}+\phi^{2}\\
  \hline
  (2,3)& 16& \phi^{2}\\
  \hline
  (2,4)& 17& \phi^{2}+\phi^{3}\\
  \hline
  (3,4)& 23& \phi^{3}\\
  \hline
  (4,1)& 17& \phi^{4}+\phi^{5}\\
  \hline
  (4,5)& 23& \phi^{4}\\
  \hline
  (5,1)& \phantom{0}7& \phi^{5}\\
  \hline
  (5,2)& 16& \phi^{5}+\phi^{1}\\
  \hline
    \end{array}
  \]
  \end{small}
 \end{minipage}
 \caption{An example of toric diagram and the chiral fields of the
 associated quiver gauge theory} \label{fig:toric-data} }

\FIGURE{
 \includegraphics[width=.33\textwidth]{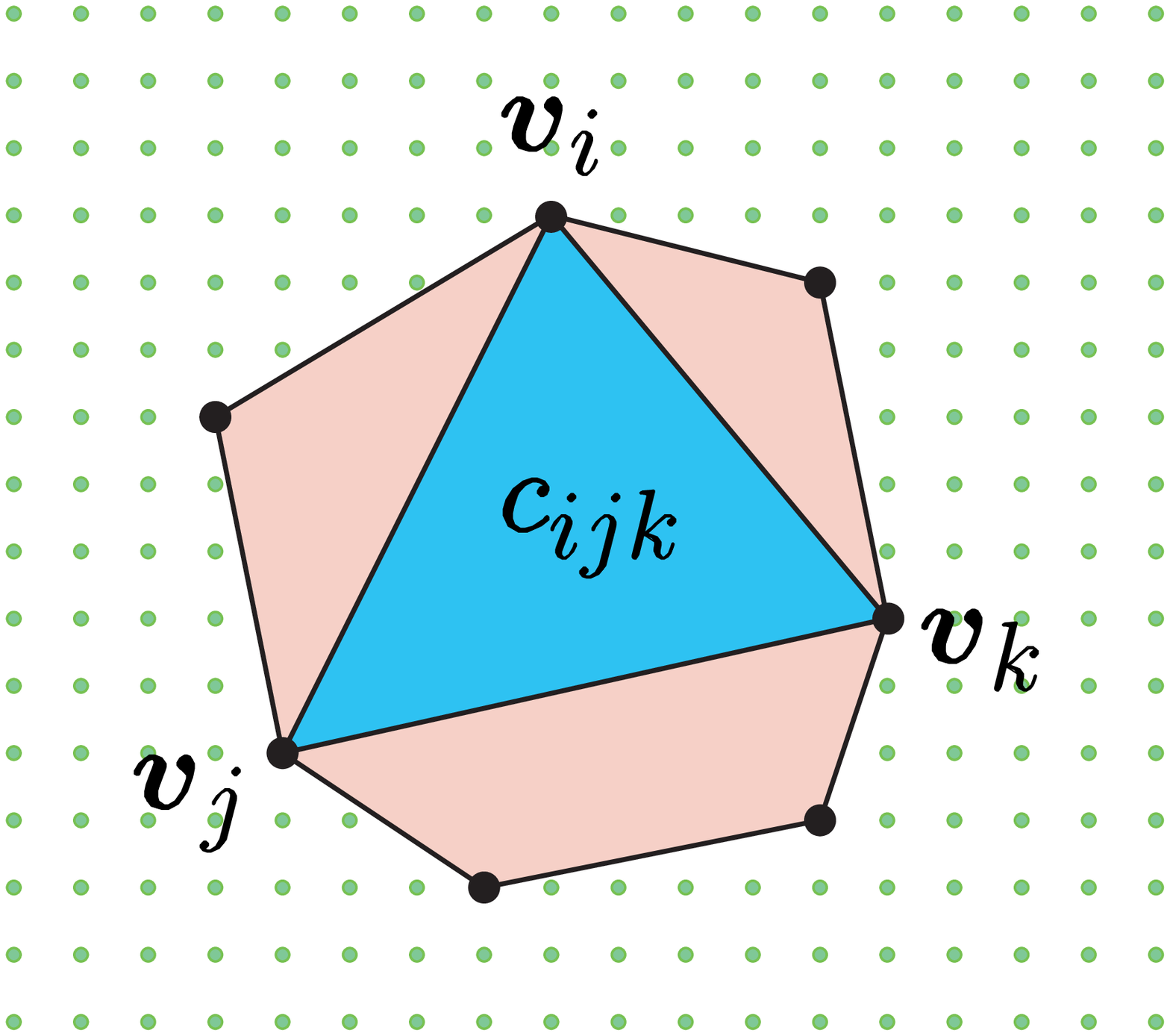}
 \caption{The coefficient $c_{ijk}$.
 \label{fig:toric-coeff}}
}
The $R$-charges of chiral bi-fundamental fields are given as follows.
Let $\mathcal{B}$ be the set of all the unordered pairs of edges of $P$. 
An element $\{\{\v{v}_{i-1},\v{v}_{i}\},\{\v{v}_{j-1},\v{v}_{j}\}\}$ of
$\mathcal{B}$ will be simply denoted by $(i,j)$, with the convention
that the oriented edge $(\v{v}_{i-1},\v{v}_{i})$ can be rotated to
$(\v{v}_{j-1},\v{v}_{j})$ in the counter-clockwise direction with an
angle $\leq 180^{\circ}$. 
For each $(i,j)\in \mathcal{B}$, we introduce
a chiral field $\Phi_{(i,j)}$ of $R$-charge 
\[
R(\Phi_{(i,j)}):=\phi^{i}+\phi^{i+1}+\dots+\phi^{j-1} 
\] 
with the
multiplicity 
\[
 \mu_{(i,j)}:=
\vev{\v{v}_{i}-\v{v}_{i-1},\v{v}_{j}-\v{v}_{j-1},\v{e}_{3}}.
\] 
By our orientation convention, $\mu_{(i,j)}$ are non-negative.  The
$R$-charges $\phi^{i}$ are constrained as
\begin{equation}
 \label{eq:Rsum=2}
 \phi^{1}+\phi^{2}+\dots+\phi^{n}=2.
\end{equation}

The $a$-function of the quiver gauge theory is then given by
\begin{equation}
 \label{eq:a-func-def-1}
  a(\phi)= \frac{9N^{2}}{32}
  \biggl[N_{\mathrm{gauge}}
  +\sum_{(i,j)\in \mathcal{B}} \mu_{(i,j)}\bigl(
  R(\Phi_{(i,j)})-1\bigr)^{3}
  \biggr].
\end{equation}
Figure \ref{fig:toric-data} is an example of a toric diagram and the
chiral field content of the corresponding quiver gauge theory.

Benvenuti, Zayas, Tachikawa \cite{Benvenuti:2006xg} and Lee, Rey
\cite{Lee:2006ru} have shown that under the constraint
(\ref{eq:Rsum=2}), the $a$-function (\ref{eq:a-func-def-1}) can be
neatly rewritten as
\begin{equation}
 \label{eq:a-def}
 a(\phi)=
  \frac{9}{32}\frac{N^{2}}{2} 
  \sum_{i,j,k=1}^{n} c_{ijk}\,\phi^{i}\phi^{j}\phi^{k} 
\end{equation}
where
\begin{equation}
 \label{eq:Cijk-def}
 c_{ijk}=|\det(\v{v}_{i},\v{v}_{j},\v{v}_{k})|
\end{equation}
is proportional to the area of the triangle with vertices
$\v{v}_{i},\v{v}_{j},\v{v}_{k}$ sitting inside $P$ (see Figure
\ref{fig:toric-coeff}). The formulas (\ref{eq:a-def}) and
(\ref{eq:Cijk-def}) make the starting point our investigation.

\section{Mathematical setup and main results}

In this section, we discuss the mathematical formulation of
$a$-maximization and state our main theorems.

Let $P\in \T$ be a toric diagram and
$\v{v}_{1},\v{v}_{2},\dots,\v{v}_{n}$ the vertices of $P$ in
counter-clockwise order, as described in Section \ref{sec:A-from-P}.
Define a homogeneous cubic polynomial $\hat{F}_{P}$ in
$\phi=(\phi^{1},\phi^{2},\dots,\phi^{n})$ by
\begin{equation}
  \label{eq:FP-def} \hat{F}_{P}(\phi) = \sum_{1\leq i<j<k\leq n}
  |\det(\v{v}_{i},\v{v}_{j},\v{v}_{k})| \, \phi^{i} \phi^{j}
  \phi^{k}.
\end{equation}
Choose a real constant $r>0$ and fix it. Let $\rho:\R_{\geq 0}^{n}\to
\R$ denote the linear function
\[
  \rho(\phi):=\phi^{1}+\dots+\phi^{n}
\]
and set
\[
   \Gamma_{n} : =\rho^{-1}(r)=\biggl\{\phi=(\phi^{1},\dots,\phi^{n})\in
   \R_{\geq 0}^{n}~: \;\sum_{i=1}^{n} \phi^{i}=r\biggr\},
\]
which is an $n{-}1$ dimensional simplex. Its \emph{relative interior}
i.e. the interior as a topological subspace of its affine hull will be
denoted by
\[
 \relint(\Gamma_{n})
=\{\phi=(\phi^{1},\dots,\phi^{n})\in \R^{n}\,:\,\sum_{i=1}^{n}
\phi^{i}=r, \;\phi^{i}> 0\}.
\]

The function $F_{P}$ is defined to be the restriction of $\hat{F}_{P}$
to $\Gamma_{n}$. Obviously, $F_{P}$ is a model for $a$-function
(\ref{eq:a-def}), and $\Gamma_{n}$ (or $\relint(\Gamma_{n})$) represents
a physically allowed region of $R$-charges. The choice of $r$ is not
important for $a$-maximization; the homogeneity
$\hat{F}_{P}(\lambda\phi)=\lambda^{3}\hat{F}_{P}(\phi)$ allows us to
choose $r$ any positive real number. Usually we set $r=2$ to match the
convention (\ref{eq:Rsum=2}).

For each toric diagram $P\in \T$, define its \emph{modulus} by
\[
 \mathfrak{M}(P):= \biggl(\frac{3}{r} \biggr)^{3}
  \max_{\phi\in \rho^{-1}(r)}\hat{F}_{P}(\phi).
\]
The modulus $\mathfrak{M}(P)$ is independent of $r$ and is normalized so
that the smallest toric diagram $P=\{\v{v}_{1}=(0,0,1),
\v{v}_{2}=(1,0,1), \v{v}_{3}=(0,1,1)\}$ has unit modulus. This
$\mathfrak{M}$ is the quantity of our primary interest.

Let $G$ be the subgroup of $GL(3,\Z)$ which leave invariant the set of
lattice points on hyperplane $(*,*,1)$. $G$ induces integral affine
transformations on this hyperplane: $G\simeq GL(2,\Z)\ltimes
\Z^{2}$. $G$ acts naturally on the set of toric diagrams $\T$: for $g\in
G$ and a polygon $P$ with vertices $\v{v}_{1},\cdots,\v{v}_{n}$, $g(P)$
is the polygon with vertices $g(\v{v}_{1}),\cdots,g(\v{v}_{n})$. The
$G$-action defines an equivalence relation $\simeq$ on $\T$:
\[
 P\simeq Q \quad 
  \stackrel{\mathrm{def}}{\Longleftrightarrow}
  \quad \exists\, g\in G \mbox{~such that~}g(P)=Q.
\]
We denote by $[P]$ the equivalence class of $P$. The functions
$\hat{F}_{P}$ are $G$-invariant, $F_{P}(\phi)$ $=F_{g(P)}(\phi)$,
because $\hat{F}_{P}$ depends on $P$ only through the areas of triangles
inscribed in $P$.  The modulus $\mathfrak{M}$ is thus well-defined on
$\T/\simeq$. In the physical context, two $G$-equivalent toric diagrams
$P$ and $Q$ are associated with the identical quiver gauge theory and
the same dual Sasaki-Einstein geometry, so there is no reason to
distinguish the two.

In connection with RG flow, it is interesting to compare
$\mathfrak{M}(P)$ and $\mathfrak{M}(P')$ for toric diagrams $P$ and $P'$
which are not necessarily $G$-equivalent. The inclusion relation
$\subset $ on $\T$ naturally induces a partial order $\preceq$ on
$\T/\simeq$, namely,
\begin{equation}
\label{eq:PartialOrder}
 [P]\preceq [P'] \;
  \stackrel{\mathrm{def}}{\Longleftrightarrow}
  \;
  \exists \, Q,Q'\in \T \mbox{~such that~}P\simeq Q,\;
  P'\simeq Q',\; Q\subset Q'.
\end{equation}

The basic question we shall be concerned with is the existence and
uniqueness of the critical point of $F_{P}$; we want to establish this
as a mathematical fact independent of duality conjectures.  This problem
is not so simple as it may appear at a first glance. For example,
consider the function
\[
  \hat{F}_{P}(\phi)=2 \phi^{1} \phi^{2} \phi^{3} +4 \phi^{1} \phi^{2}
  \phi^{4}+3 \phi^{1} \phi^{3} \phi^{4} +\phi^{2} \phi^{3}
  \phi^{4}
\]
which corresponds to the toric diagram
$P=\{(2,3,1),(0,1,1),(0,0,1),(1,0,1)\}$. This $\hat{F}_{P}$ has a unique
critical point in $\relint(\Gamma_{4})$. However,
\[
  \hat{F}(\phi)=2 \phi^{1} \phi^{2} \phi^{3} +8 \phi^{1} \phi^{2}
  \phi^{4}+3 \phi^{1} \phi^{3} \phi^{4} +\phi^{2} \phi^{3}
  \phi^{4}\qquad\qquad\quad~
\]
has no critical points in $\relint(\Gamma_{4})$; it is maximized at
$\phi=(\frac{r}{3},\frac{r}{3},0,\frac{r}{3})\in \partial\Gamma_{4}$ but
this is not a critical point. On the other hand,
\begin{eqnarray*}
  \hat{F}(\phi)&=&\, 4 \phi^{1} \phi^{2} \phi^{3}+2 \phi^{1} \phi^{2} \phi^{4}
  +9 \phi^{1} \phi^{3} \phi^{4}+7 \phi^{2} \phi^{3} \phi^{4}
  \\
  && +4 \phi^{1} \phi^{3} \phi^{5}+\phi^{2} \phi^{3} \phi^{5}
  +\phi^{1} \phi^{4} \phi^{5}+10 \phi^{2} \phi^{4} \phi^{5}
  +4 \phi^{3} \phi^{4} \phi^{5}
\end{eqnarray*}
has two critical points in $\relint(\Gamma_{5})$; one is a local maximum
and the other is a saddle point. Consequently
the conditions such as 
\begin{itemize}
 \itemsep=-3pt
 \item  $c_{ijk}$ are non-negative integers,
 \item  $c_{ijk}$ is invariant under any permutation of indices
      $i,j,k$, and
 \item  $c_{ijk}=0$ unless $i,j,k$ are distinct,
\end{itemize}
are not enough to guarantee the existence and uniqueness of the critical
point in $\relint(\Gamma_{n})$. The fact that the coefficients $c_{ijk}$
are given by the areas of triangles will be heavily used in this paper.

Our main results are as follows\footnote{The integrality of vertices
$\v{v}_{1},\dots,\v{v}_{n}$ will be important for constructing quiver
gauge theories or Calabi-Yau cones. But the integrality is not needed to
establish the analytic properties of $\hat{F}_{P}$ obtained in this
paper. In fact, $P$ can be any convex polygon on a hyperplane not
passing through the origin.}:
\begin{theorem}[Theorem \ref{thm:main1-again}]
 \label{thm:main1} The function $F_{P}:\Gamma_{n}\to \R$ has a unique
 critical point $\phi_{*}$ in $\relint(\Gamma_{n})$ and $\phi_{*}$ is
 also the unique global maximum of $F_{P}$.
\end{theorem}

\begin{theorem}[Theorem \ref{thm:PhysicalBound}]
 \label{thm:main2} The critical point $\phi_{*}$ satisfies
 the universal bound 
 \[
    0<\phi_{*}^{i}\leq \frac{r}{3}\qquad (i=1,\dots,n).
 \]
 Here, the equality $\phi_{*}^{i}=\frac{r}{3}$ holds for some $i$ if and
only if $n=3$.
\end{theorem}

\begin{theorem}[Theorem \ref{thm:Monotonicity2}]
 \label{thm:main3} The maximum value of $F_{P}$ is monotone in the
 following sense: Suppose $P$ and $P'$ are toric diagrams satisfying
 $[P]\preceq[P']$. Then $\mathfrak{M}(P)\leq \mathfrak{M}(P')$. The
 equality holds if and only if $P\simeq P'$.
\end{theorem}

Some comments are in order here. The unitarity of the representation of
superconformal algebra $SU(2,2|1)$ requires that all gauge invariant
chiral operators must have $U(1)_{R}$ charge $R\geq \frac{2}{3}$
\cite{MR812225,MR740818}. Theorem \ref{thm:main2}, however, yields 
opposite inequalities $\phi^{i}_{*}\leq \frac{2}{3}$ in the conventional
normalization $r=2$ (see (\ref{eq:Rsum=2})). This is not a contradiction
because $\phi_{*}^{i}$ are $R$ charges of gauge non-invariant
bi-fundamental fields.

Theorem \ref{thm:main3} can be regarded as a combinatorial analogue of
``$a$-theorem'': the $a$-function always decreases whenever the toric
diagram shrinks.

There are two key ingredients in the proof of the main results. First,
$a$-function is identified with the volume of a three dimensional
polytope called ``zonotope'' (Proposition \ref{prop:a=vol(Z)});
Brunn-Minkowski inequality asserts that (cubic root of) volume function
is a concave function on the space of polytopes. This concavity
guarantees the uniqueness of the critical point (Proposition
\ref{prop:CriticalIsUnique}). Second key point is to show the
monotonicity of modulus $\mathfrak{M}$ under simple change of the toric
diagrams, e.g. deleting a vertex. This property is also used to prove
the existence of the critical point.

Here is an application of our results. The uniqueness of the maximizer
implies that there is no spontaneous symmetry break down in
$a$-maximization:
\begin{corollary}
 If a nontrivial element $g$ of $G$ fixes a toric diagram $P$, then the
 critical point $\phi_{*}$ of $F_{P}$ is also fixed by $g$.
\end{corollary}

\section{Polytopes and Zonotopes}

Let $\R^{d}$ denote a $d$-dimensional real vector space. A subset
$C\subset \R^{d}$ is called \textit{convex} if 
$(1-\lambda)\v{x}+\lambda \v{y}\in
C $ whenever $\v{x},\v{y}\in C$ and $0\leq \lambda\leq 1$. For any set
$S\subset \R^d$, its convex hull $\conv(S)$ is, by definition, the
smallest convex set containing $S$:
\[
  \conv(S) := \{  \lambda\v{x}+(1-\lambda)\v{y} \in \R^{d}~:~
  \v{x}, \v{y}\in S,\;  0\leq \lambda\leq 1  \}.
\]
A \emph{polytope} is the convex hull $\conv(S)$ of a finite
set $S$ in $\R^{d}$.

\FIGURE{\epsfig{file=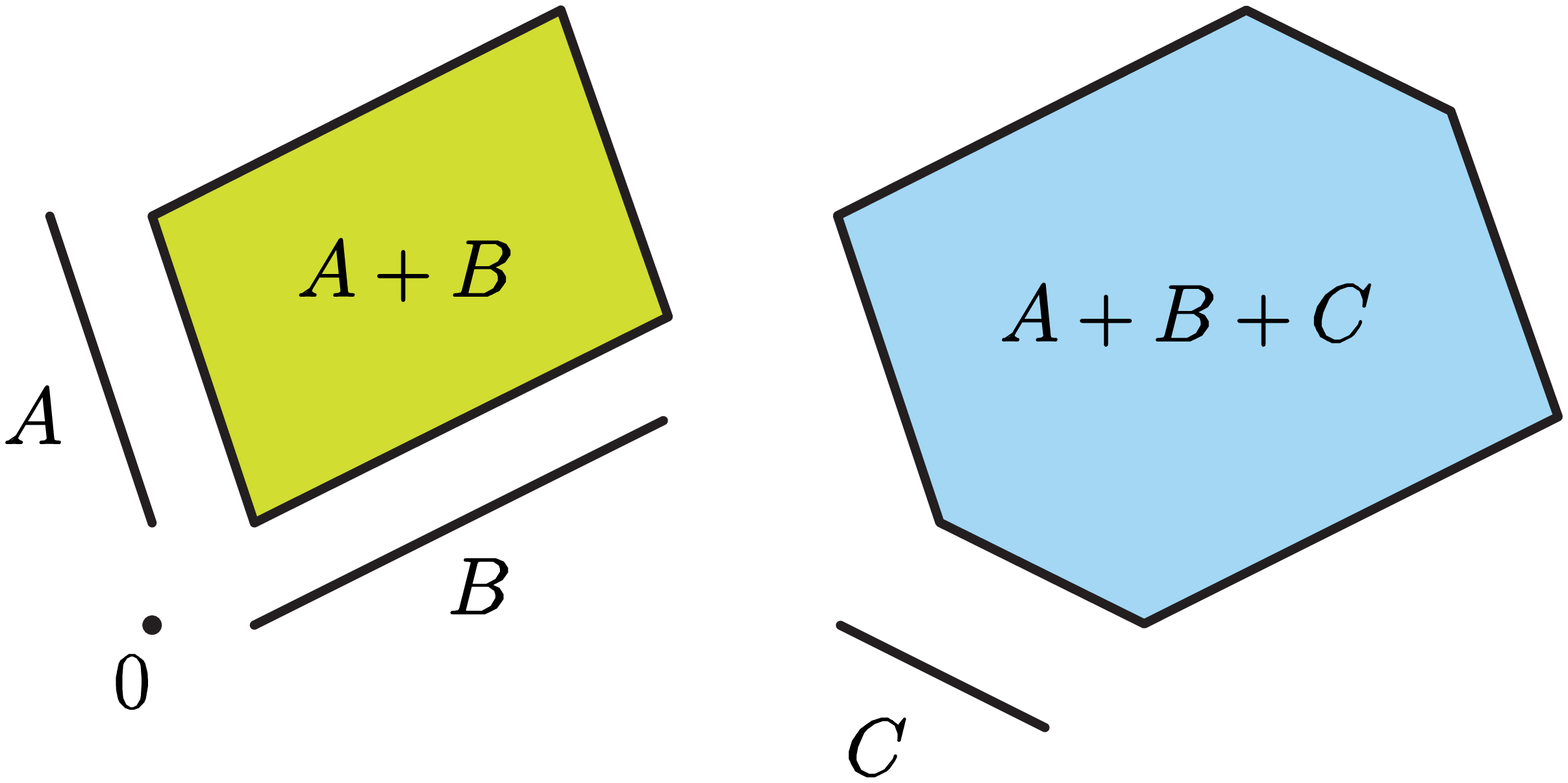,width=0.45\textwidth}
 \caption{Minkowski sums \label{fig:Minkowski-sums}}
}
The \emph{Minkowski sum}, or \emph{vector sum}, of two subsets $A$
and $B$ in $\R^{d}$ is (see Figure \ref{fig:Minkowski-sums})
\[
  A+B:= \{\v{x}+\v{y}~:~ \v{x}\in A,~\v{y}\in B\}, 
\]
whereas the \emph{dilatation} by
the factor $r\geq 0$ is
\[
  r A= \{ r \v{x} ~:~ \v{x}\in A\}.
\]
If $A$ and $B$ are polytopes, then $A+B$, $rA$ are also polytopes.  Let
$\mathcal{P}^{d}$ denote the family of all convex polytopes in $\R^{d}$. 
Two basic operations, Minkowski sum and dilatation, make the family
$\mathcal{P}^{d}$ a convex set: for any $A,B\in \mathcal{P}^{d}$ and
non-negative numbers $\alpha,\beta$ such that $\alpha+\beta=1$ one has
$\alpha A+\beta B\in \mathcal{P}^{d}$.

\FIGURE[b]{
 \epsfig{file=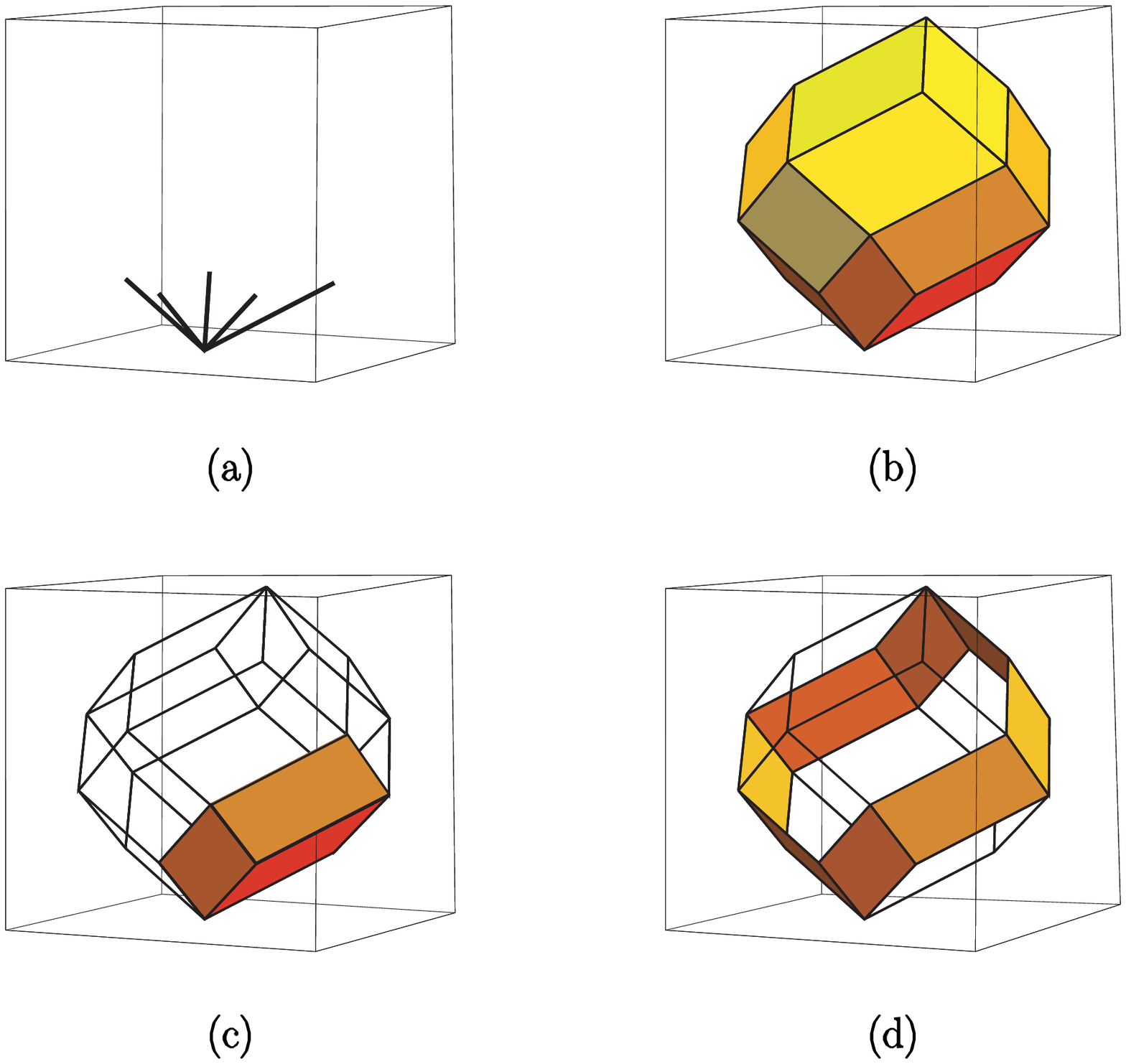,width=.60\textwidth}
 \caption{A zonotope ($d=3$, $n=5$) : generators (a), zonotope (b),
 cube (c) and zone (d).
 \label{fig:Zonotope}}
}
Let $S_{1},\dots,S_{n}$ be $n$ line segments, each of non-zero length,
in $\R^{d}$. The polytope $\cZ$ defined as the Minkowski sum
\[
  \cZ=S_{1}+\dots+S_{n}
\]
is called a \emph{zonotope} and $S_{1},\dots,S_{n}$ are called its
\emph{generators} (Figure \ref{fig:Zonotope}). For a finite collection
of vectors $X=\{\v{x}_{1},\dots,\v{x}_{n}\}\subset \R^{d}$, we put $
S_{i}=\conv(\v{0},\v{x}_{i})\; (i=1,\dots,n)$, and write $\cZ[X]$ the
corresponding zonotope. Equivalently,
\[
  \cZ[X]=\{\v{x}\in \R^{d}~:~ \v{x}=
 \lambda_{1}\v{x}_{1}+\dots+ \lambda_{n}\v{x}_{n},
 ~0\leq \lambda_{i}\leq 1,~i=1,\dots,n
 \}.
\]
The zonotope $\cZ[X]$ is the image of an $n$-dimensional cube $[0,1]^{n}$
under a linear projection $\pi:\R^{n}\to \R^{d}$ defined by the $d\times
n$ matrix $X$.  $\cZ[X]$ may also be defined as the convex hull of
$2^{n}$ points
\[
 \{ \v{x}_{i_{1}}+\v{x}_{i_{2}}+\dots +\v{x}_{i_{k}}\in \R^{d}~:~
  1\leq i_{1}<i_{2}<\dots<i_{k}\leq n,~ k=0,1,\dots,n\}.
\]
$\cZ[X]$ is centrally symmetric, and its center is located at
$\frac{1}{2}(\v{x}_{1}+\dots+\v{x}_{n})$.

The zonotope $\cZ[X]$ can be decomposed into {\scriptsize$\pmatrix{n\cr
d}$} $d$-dimensional parallelepipeds called \emph{cubes}, each of which
is a translation of
\[
  Q_{i_{1},\dots,i_{d}}:=\conv(\v{0},\v{x}_{i_{1}})+
  \conv(\v{0},\v{x}_{i_{2}})+
  \dots+\conv(\v{0},\v{x}_{i_{d}}).
\]
The crucial fact is that, although such decomposition is not unique, all
$d$-tuples $\{\v{x}_{i_{1}},$ $\dots,$ $\v{x}_{i_{d}}\}$ $\subset X$ appear
exactly once in any decomposition. Since the volume of each cube
$Q_{i_{1},\dots,i_{d}}$ is simply given by
$\vol(Q_{i_{1},\dots,i_{d}})=|\det(\v{x}_{i_{1}},\dots,\v{x}_{i_{d}})|$,
this leads to the following volume formula for zonotopes, which will
play a crucial role in this paper.
\begin{theorem}[Shephard  \cite{MR0362054}, attributed to McMullen]~
\begin{equation}
 \label{eq:volZ}
 \vol(\cZ[X]) = \sum_{1\leq i_{1}<\dots<i_{d}\leq n}
  |\det(\v{x}_{i_{1}},\dots,\v{x}_{i_{d}})|.
\end{equation}
\end{theorem}

In the rest of this paper, we will specialize to $d{=}3$ case, i.e. three
dimensional zonotopes.  If no three of the $n$ vectors
$\v{x}_{1},\dots,\v{x}_{n}$ are coplanar, all the facets (i.e. two
dimensional faces) of $\cZ[X]$ are parallelograms. For a given generator
$\v{x}_{s}$, the faces which has a edge parallel to $\v{x}_{s}$ form a
\emph{zone} going around a zonotope. Each zone consists of $n{-}1$ pairs
of opposite faces, there are altogether $n(n{-}1)/2$ pairs of opposite
faces, $n(n{-}1)$ pairs of opposite edges, and therefore $n(n{-}1)/2+1$
pairs of opposite vertices.

The next Proposition is our key observation, which immediately follows
by comparing (\ref{eq:FP-def}) and (\ref{eq:volZ}).
\begin{proposition}
 \label{prop:a=vol(Z)} Let $P\in \T$ be a toric diagram with vertices
 $\v{v}_{1},\dots,\v{v}_{n}$. The function $\hat{F}_{P}(\phi)$ defined
 in (\ref{eq:FP-def}) is equal to the volume of the zonotope
 \begin{equation}
 \label{eq:ZonotopeFamily}
 \cZ_{P}(\phi):= 
  \phi^{1}\conv(\v{0},\v{v}_{1})+ \phi^{2}
  \conv(\v{0},\v{v}_{2})
  +\dots +\phi^{n}\conv(\v{0},\v{v}_{n}).
 \end{equation}
 The trial $a$-function (\ref{eq:a-def}) is therefore given by
 \[
    a(\phi)=\frac{27}{32}N^{2} \hat{F}_{P}(\phi)
  =\frac{27}{32}N^{2} \vol(\cZ_{P}(\phi)).
 \]
\end{proposition}

\section{Uniqueness of the critical point}

In this section, the uniqueness of the critical point of $ \hat{F}_{P}$
is proved; the existence is shown in the next section. 

A real-valued function $f$ on a convex set $C$ is \emph{concave} if
\[
  f((1-\lambda)x + \lambda y) \geq (1-\lambda)f(x) +\lambda f(y)
\]
for all $x, y \in C$ and $0 <\lambda <1$. If the above inequality can be
replaced by
\[
  f((1-\lambda)x + \lambda y) > (1-\lambda)f(x) +\lambda f(y),
\]
then $f$ is \emph{strictly concave}. We will use the following well
known properties of concave functions:
\begin{theorem}
 \label{thm:ConcaveFunctions} Any local maximizer of a concave function
 $f$ defined on a convex set $C$ of $\R^{n}$ is also a global maximizer
 of $f$. If in addition $f$ is differentiable, then any
 stationary point is a global maximizer of $f$.
 Any local maximizer of a strictly concave function $f$ defined
 on a convex set $C$ of $\R^{n}$ is the unique strict global maximizer
 of $f$ on $C$.  
\end{theorem}

Recall that the family $\mathcal{P}^{d}$ of polytopes in $\R^{d}$ is a
convex set under the operations Minkowski sum and dilatation.  Thus it
makes sense to talk about the concavity of a function defined on
$\mathcal{P}^{d}$, such as volume function.  The following is a
fundamental result in the theory of convex bodies (for extensive survey,
see \cite{MR1216521,MR1898210}).

\begin{theorem}[Brunn-Minkowski inequality]
 \label{eq:BMinequality} The $d$-th root of volume is a concave
 function on the family of convex bodies in $\R^{d}$. More
 precisely, for convex bodies $A, B\subset \R^{d}$ and for $0\leq
 \lambda \leq 1$,
 \[
    \left(\vol((1-\lambda)A+\lambda B)\right)^{1/d}
   \geq 
  (1-\lambda)\left(\vol(A)\right)^{1/d}
  +  \lambda\left(\vol(B)\right)^{1/d}.
 \]
 Equality for some $0<\lambda<1$ holds if and only if $A$ and $B$ either
 lie in parallel hyperplanes or are homothetic. \footnote{Two sets $A,
 B\subset \R^{n}$ are called \emph{homothetic} if $A=\kappa B + t$ for
 some $\kappa> 0$ and $t\in \R^{n}$, or one of them is a single point.}
\end{theorem}

In Proposition \ref{prop:a=vol(Z)}, $F_{P}(\phi)$ is identified with the
volume of a three dimensional zonotope. Actually, we are interested in
the ``family'' of zonotopes $\cZ_{P}(\phi)$ parametrized by
$\phi=(\phi^{1},\dots,\phi^{n})\in \Gamma_{n}$. In order to apply
Brunn-Minkowski inequality to this family, let us investigate under what
conditions two zonotopes are homothetic to each other.
\begin{lemma}
 \label{lem:homothety} For $\phi,\phi'\in \R_{\geq 0}^{n}$, two
 zonotopes $\cZ_{P}(\phi)$, $\cZ_{P}(\phi')$ of nonzero volume are
 homothetic if and only if $\phi=\kappa \phi'$ for some $\kappa>0$. In
 particular, two zonotopes $\cZ_{P}(\phi)$, $\cZ_{P}(\phi')$ with
 $\phi,\phi'\in \Gamma_{n}$ are homothetic if and only if $\phi=\phi'$.
\end{lemma}
\Proof
 Suppose $\cZ_{P}(\phi)$ and $\cZ_{P}(\phi')$ are homothetic. By
 assumption, they have nonzero volume and cannot be in a
 hyperplane. Thus there exists $\kappa>0$ and $t\in \R^{3}$ such that
 $\cZ_{P}(a)=\kappa \cZ_{P}(b)+t$. In fact $t=0$ because both zonotopes
 have $O=(0,0,0)$ as the bottom vertex. Each of them have a unique edge
 starting from $O$ and parallel to $\v{v}_{i}$ for all $i$. Homothethy
 implies $\phi^{i}\conv(\v{0},\v{v}_{i})=\kappa
 \phi'^{i}\conv(\v{0},\v{v}_{i})$, so $\phi^{i}=\kappa \phi'^{i}$ holds
 for all $i$. In particular, if $\phi,\phi'\in \Gamma_{n}$, then
 $r=\rho(\phi)=\rho(\kappa\phi')=\kappa \rho(\phi')=\kappa r$, so
 $\kappa=1$.  \qed

Here we come to the key point of our analysis.
\begin{proposition}
 \label{prop:fP-concavity} The function 
 \begin{equation}
  \label{eq:def-of-f}
   \bigl(\hat{F}_{P}(\phi)\bigr)^{1/3}
   =\left(\vol \cZ_{P}(\phi)\right)^{1/3} ~:~\R_{\geq 0}^{n}\to \R
 \end{equation}
 is concave. Moreover, its restriction, $(F_{P})^{1/3} : \Gamma_{n}\to
 \R$ is strictly concave.
\end{proposition}
\Proof
 Let us denote the function (\ref{eq:def-of-f}) by $f_{P}$. It suffices
 to show that for any $a=(a^{1},\dots,a^{n})$, $b=(b^{1},\dots,b^{n})$
 $\in \R_{\geq 0}^{n}$,
 \[
    f_{P} ((1-\lambda)a + \lambda b) \geq (1-\lambda)f_{P}( a) 
   + \lambda f_{P}(b)
   \qquad (0\leq \lambda\leq 1)
 \]
 and the equality holds if and only if $a=\kappa\, b$ for some
 $\kappa>0$. One can easily check that
 \[
    \sum_{i=1}^{n}((1-\lambda)a^{i} + \lambda b^{i})\conv(\v{0},\v{v}_{i})
   =
   (1-\lambda)  \sum_{i=1}^{n}a^{i}\conv(\v{0},\v{v}_{i})
   +
   \lambda \sum_{i=1}^{n}b^{i}\conv(\v{0},\v{v}_{i})
 \]
 holds as an equality in $\mathcal{P}^{d}$. Using the notation
 (\ref{eq:ZonotopeFamily}), this is written as
 \[
    \cZ_{P}( (1-\lambda)a + \lambda b)=(1-\lambda)\cZ_{P}(a)+\lambda \cZ_{P}(b)
 \]
 Then the claim immediately follows from Theorem \ref{eq:BMinequality}
 and Lemma \ref{lem:homothety}. 
 \qed

 Since the function $x \mapsto x^{1/3}:\R_{\geq 0}\to \R_{\geq 0}$ is a
 strictly increasing function, $\hat{F}_{P}$ is maximal (resp. critical)
 at $\phi$ if and only if $(\hat{F}_{P})^{1/3}$ is maximal
 (resp. critical) at $\phi$. Combining Theorem
 \ref{thm:ConcaveFunctions} and Proposition \ref{prop:fP-concavity}, we
 have established the uniqueness of the solution to $a$-maximization:
 \begin{proposition}
  \label{prop:CriticalIsUnique} Suppose $\phi_{*}$ is a critical point
  or a local maximum of $F_{P}:\Gamma_{n} \to \R$. Then $\phi_{*}$ is
  the unique critical point and is also the global maximum over
  $\Gamma_{n}$.
 \end{proposition}

 A remark is in order here: 
 $F_{P}$ is not necessarily concave although
 the cubic root $(F_{P})^{1/3}$ is. The conifold $
 \v{v}_{1}= (1,1,1) , \v{v}_{2}= (1,0,1) , \v{v}_{3}= (1,0,0) ,
 \v{v}_{4}= (1,1,0)$, 
 \[
    F_{P}(\phi)=\phi^{2}\phi^{3}\phi^{4}+
  \phi^{1}\phi^{3}\phi^{4}+\phi^{1}\phi^{2}\phi^{4}+\phi^{1}\phi^{2}\phi^{3},
  \qquad (\phi^{1}+\phi^{2}+\phi^{3}+\phi^{4}=2)
 \]
 is already a counterexample; the Hessian of $F_{P}$ at
 $\phi=(\frac{1}{7},\frac{1}{7},\frac{3}{7},\frac{9}{7})$ is not
 negative definite.

\section{Existence of the critical point}

This section is devoted to the proof of Theorem \ref{thm:main1} (Theorem
\ref{thm:main1-again}). The key idea is as follows. The continuous
function $F_{P}$ always has a global maximum on the closed set
$\Gamma_{n}$. If the maximum point is in $\relint(\Gamma_{n})$, then
from Proposition \ref{prop:CriticalIsUnique} it is also a critical point
and there is no other local maxima.  But if the maximum point is on the
boundary $\partial \Gamma_{n}$, it is not necessarily a critical point ---
physical SCFT. Therefore to establish Theorem \ref{thm:main1}, it
suffices to show that a point on the boundary $\partial \Gamma_{n}$ can
never be a local maximum of $F_{P}$.

For this purpose, we investigate the behavior of the maximum values
under the change of toric diagrams. More precisely we will prove the
following
\begin{proposition}
 \label{prop:Monotonicity} Let $P$ be a toric diagram with vertices
 $\v{v}_{1},\dots,\v{v}_{n}$ in counter\-clock\-wise order. Let $Q$ be a
 toric diagram obtained by deleting $\v{v}_{n}$ from $P$, i.e. the
 convex hull of $n{-}1$ vertices $\v{v}_{1},\dots,\v{v}_{n-1}$ as in
 Figure \ref{fig:enlargingP}. Then, for any $\phi\in
 \relint(\Gamma_{n-1})$, there exits $\psi\in \relint(\Gamma_{n})$ such
 that $F_{Q}(\phi)<F_{P}(\psi)$.
\end{proposition}

\FIGURE{
 \epsfig{file=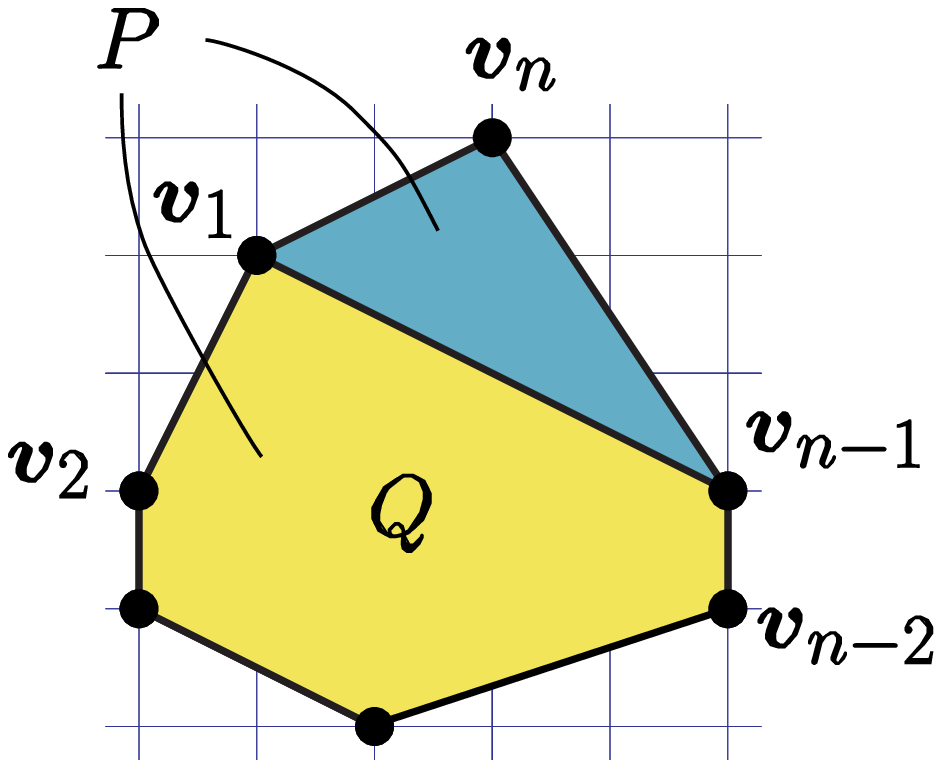,width=.28\textwidth}
 \caption{Deleting a vertex
 \label{fig:enlargingP}}
}

Note that under the natural inclusion 
\[
  \Gamma_{n-1}\subset \Gamma_{n},\qquad 
  \phi=(\phi^{1},\dots,\phi^{n-1})\mapsto
  (\phi^{1},\dots,\phi^{n-1},0),
\]
the point $\phi\in \relint (\Gamma_{n-1})$ corresponds to a point on the
boundary facet $\{\phi^{n}=0\}$ of $\Gamma_{n}$, and $F_{Q}$ is none
other than the restriction of $F_{P}$ to this facet. Clearly, any
boundary point of $\Gamma_{n}$ is obtained in this manner. Thus Proposition
\ref{prop:Monotonicity} immediately implies
\begin{corollary}
 \label{cor:MaxInterior}
 No boundary point of $\Gamma_{n}$ can be a local maximum of
 $F_{P}:\Gamma_{n}\to \R$.
\end{corollary}

\begin{corollary}
 \label{cor:Monotonicity} Suppose a toric diagram $Q$ is obtained from
 another toric diagram $P$ by removing one vertex, then
 $\mathfrak{M}(Q)<\mathfrak{M}(P)$.
\end{corollary}

By the argument given in the first paragraph of this section, we deduce
from Corollary \ref{cor:MaxInterior} the following
\begin{theorem}
 \label{thm:main1-again} Suppose $P$ is a toric diagram with vertices
 $\v{v}_{1},\dots,\v{v}_{n}$. Then $F_{P}:\Gamma_{n}\to \R$ has a unique
 critical point $\phi_{*}$ in $\relint(\Gamma_{n})$ and $\phi_{*}$ is
 also the unique global maximum of $F_{P}$.
\end{theorem}

Let us turn to the proof of Proposition \ref{prop:Monotonicity}. Our
strategy is to show that for any $\phi\in \relint(\Gamma_{n-1})\subset
\partial \Gamma_{n}$ there is at least one ``inward'' direction in which
$F_{P}$ is strictly increasing. Consider two straight paths
$\psi_{I}(t), \psi_{II}(t)$ in $\Gamma_{n}$ emanating from the boundary
point $\phi$, defined by
\[
   \psi_{I}^{i}(t)=
\left\{
   \begin{array}{ll}
    \phi^{1}-t
     \quad & \mbox{if $i=1$},\\
    t&\mbox{if $i=n$},\\
    \phi^{i}&\mbox{otherwise},\\
   \end{array}
\right.
   \qquad \mbox{and} \qquad 
 \psi_{II}^{i}(t)=
\left\{
 \begin{array}{ll}
  \phi^{n-1}-t
   \quad& \mbox{if $i=n-1$},\\
  t& \mbox{if $i=n$},
  \\
  \phi^{i}&\mbox{otherwise},
 \end{array}
\right.
\]
for $0\leq t\leq \min(\phi^{1},\phi^{n-1})$.  
If either $\frac{d}{dt}\big|_{t=0}
F_{P}(\psi_{I}(t))>0$ or $\frac{d}{dt}\big|_{t=0}
F_{P}(\psi_{II}(t))>0$ is proved, then we are done.

Note that for three vectors $\v{a},\v{b},\v{c}\in \R^{3}$, the relation
$\vev{\v{a},\v{b},\v{c}}=(\v{a}\times \v{b})\cdot \v{c}$ holds, where
$\v{a}\times\v{b}$ denotes the cross product and $\cdot$ is the standard
inner product. Let $\v{\xi}\in \R^{3}$ be a vector defined by
\[
   \v{\xi}=\sum_{1\leq i<j\leq n-1} \phi^{i} \phi^{j}
   \v{v}_{i}\times \v{v}_{j}.
\]
It is easy to see
\begin{eqnarray*}
   \frac{d}{dt}\Big|_{t=0}  F_{P}(\psi_{I}(t))
   &
   =& \sum_{1\leq i<j\leq n-1}(c_{ij \,n}-c_{ij \,1}) \phi^{i}\phi^{j}
   = \v{\xi}\cdot (\v{v}_{n}-\v{v}_{1}),
   \\
   \frac{d}{dt}\Big|_{t=0}  F_{P}(\psi_{II}(t))
   &
   =& \sum_{1\leq i<j\leq n-1}(c_{ij \,n}-c_{ij \,n{-}1}) \phi^{i}\phi^{j}
   =  \v{\xi}\cdot (\v{v}_{n}-\v{v}_{n-1}).
\end{eqnarray*}
Thus it suffices to show either $\v{\xi}\cdot (\v{v}_{n}-\v{v}_{1})>0$ or
$ \v{\xi}\cdot (\v{v}_{n}-\v{v}_{n-1})>0$ holds.  

\FIGURE{ \epsfig{file=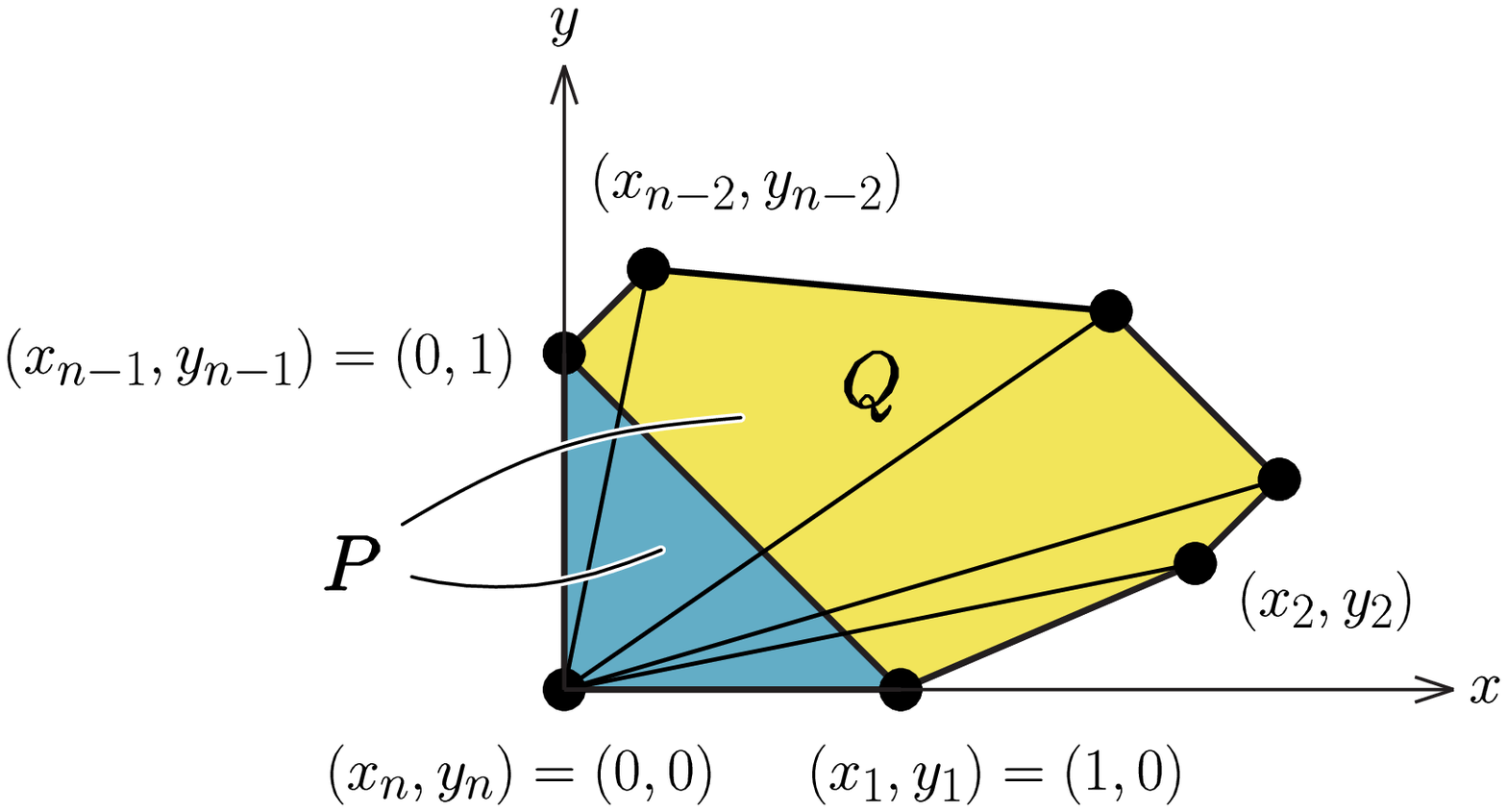,width=.45\textwidth}
 \caption{$(x,y)$-coordinates.
 \label{fig:xy-coordinates}}
}
Let us choose three vectors
$\v{v}_{1},\v{v}_{n-1},\v{v}_{n}$ as a basis of $\R^{3}$ and 
express other $\v{v}_{i}$'s as
\[
  \v{v}_{i} = x_{i}\v{v}_{1} + y_{i}\v{v}_{n-1} +(1-x_{i}-y_{i})\v{v}_{n}.   
\]
In the affine coordinates $(x_{i},y_{i})$, the toric diagram $P$ and $Q$
looks like polygons sitting in the first quadrant of $\R^{2}$, as
depicted in Figure \ref{fig:xy-coordinates}. 
Note that $x_{i}y_{j}>x_{j}y_{i}$ for all $1\leq i<j\leq n-1$.

A straightforward calculation shows 
\begin{eqnarray*}
  \v{\xi}\cdot (\v{v}_{n}-\v{v}_{1}) &= &
  \det(\v{v}_{1},\v{v}_{n-1},\v{v}_{n})
  \sum_{1\leq i<j\leq n-1} \phi^{i} \phi^{j}
  (y_{j}-y_{i}),
  \\
  \v{\xi}\cdot (\v{v}_{n}-\v{v}_{n-1}) &=& 
  \det(\v{v}_{1},\v{v}_{n-1},\v{v}_{n})
  \sum_{1\leq i<j\leq n-1} \phi^{i} \phi^{j}
  (x_{i}-x_{j}). 
\end{eqnarray*}
Since $\det(\v{v}_{1},\v{v}_{n-1},\v{v}_{n})>0$, the claim follows from
the next lemma.
\begin{lemma}
 Let $\phi^{1},\phi^{2},\dots,\phi^{n-1}$ be positive
 numbers and 
 \[
    (x_{1},y_{1}), \;  (x_{2},y_{2}),\;  \dots\;  ,
   (x_{n-2},y_{n-2}),\;  (x_{n-1},y_{n-1})
 \]
 be $n-1$ points in $\R_{\geq 0}^{2}$ satisfying
  \begin{equation}
   \label{eq:ArgumentIncreases}
   x_{i}y_{j}-x_{j}y_{i}>0 \qquad (1\leq i<j\leq n-1).
  \end{equation}
 Then, at least one of the following inequalities holds:
 \begin{equation}
  \label{eq:aaa}
  \sum_{1\leq i<j\leq n-1}\phi^{i}\phi^{j}(x_{i}-x_{j})>0,
  \qquad \mbox{or}
  \quad \sum_{1\leq i<j\leq n-1}\phi^{i}\phi^{j}(y_{j}-y_{i})>0.
 \end{equation}
\end{lemma}
\Proof
 Define $c_{1},c_{2},\dots,c_{n-1}$ by
 \[
    c_{i}=\phi^{i}
   \biggl(
    \sum_{1\leq j<i}\phi^{j}
    -\sum_{i<j\leq n-1}\phi^{j}
   \biggr).
 \]
 Clearly $c_{1}<0$ and $c_{n-1}>0$, and the sequence
 $\frac{c_{1}}{\phi^{1}},\frac{c_{2}}{\phi^{2}},
 \dots,\frac{c_{n-1}}{\phi^{n-1}}$ is strictly increasing since
 $\frac{c_{i+1}}{\phi^{i+1}}-\frac{c_{i}}{\phi^{i}}$
 $=\phi^{i}+\phi^{i+1}>0$. Hence there exists $k$ ($2\leq k\leq n-1$)
 such that $c_{1},c_{2},\dots,c_{k-1}<0$ and $c_{k},c_{k+1},\dots ,
 c_{n-1}\geq 0$.  The expressions is (\ref{eq:aaa}) can be rewritten as
 \begin{eqnarray}
   \sum_{1\leq i<j\leq n-1}\phi^{i}\phi^{j}(x_{i}-x_{j})
   &=&-\sum_{i=1}^{n-1} c_{i} x_{i}= 
   \sum_{i=1}^{k-1} |c_{i}| x_{i} -\sum_{i=k}^{n-1} |c_{i}|x_{i},
   \label{eq:aaa1a}
   \\
   \sum_{1\leq i<j\leq n-1}\phi^{i}\phi^{j}(y_{j}-y_{i})
   &=& \sum_{i=1}^{n-1} c_{i} y_{i}= 
   -\sum_{i=1}^{k-1} |c_{i}| y_{i} +\sum_{i=k}^{n-1} |c_{i}|y_{i}.
  \label{eq:aaa1b}
 \end{eqnarray}
 Suppose that, contrary to our claim, neither of (\ref{eq:aaa}) is true.
 Then (\ref{eq:aaa1a}) and (\ref{eq:aaa1b})  are both non-positive, which
 means
 \begin{equation}
  \label{eq:bbb}
  0< \sum_{i=1}^{k-1} |c_{i}| x_{i} \leq  \sum_{i=k}^{n-1} |c_{i}|x_{i}
  \qquad    \mbox{and}\qquad 
  0< \sum_{i=k}^{n-1} |c_{i}|y_{i} \leq   \sum_{i=1}^{k-1} |c_{i}| y_{i}.
 \end{equation}
 On the other hand, from (\ref{eq:ArgumentIncreases}),
 \begin{equation}
  \label{eq:ccc}
  \sum_{i=1}^{k-1} |c_{i}| (x_{i}y_{k-1}-x_{k-1}y_{i})\geq 0,
   \qquad    \mbox{and}\qquad 
  \sum_{i=k}^{n-1} |c_{i}| (x_{i}y_{k}-x_{k}y_{i})\leq 0.
 \end{equation}
 It follows from (\ref{eq:bbb}) and (\ref{eq:ccc}) that
 \[
    \frac{y_{k}}{x_{k}}\leq 
   \frac{\sum_{i=k}^{n-1} |c_{i}|y_{i}}{\sum_{i=k}^{n-1} |c_{i}|x_{i}}
   \leq
   \frac{\sum_{i=1}^{k-1} |c_{i}| y_{i}}{\sum_{i=1}^{k-1} |c_{i}| x_{i} }
   \leq   \frac{y_{k-1}}{x_{k-1}},
 \]
 therefore $x_{k-1}y_{k}-x_{k}y_{k-1}\leq 0$. This contradicts
 (\ref{eq:ArgumentIncreases}). \qed

\section{Bounds on critical points}

In this section, we establish Theorem \ref{thm:main2} (Theorem
\ref{thm:PhysicalBound}). As we will see, an upper bound on the
coordinates of the critical point is easily obtained using the
interpretation as volumes.

Let $P$ be a toric diagram with vertices $\v{v}_{1},\dots,\v{v}_{n}$.
As mentioned before, the zonotope $\cZ_{P}(\phi)$ can be cut into the
union of {\scriptsize$\pmatrix{n\cr 3}$} cubes and
$F_{P}(\phi)=\vol(\cZ_{P}(\phi))$ equals the sum of the volumes of all
cubes. We arbitrarily fix such a decomposition. For any $s$
($s=1,\dots,n$), let $\cZ_{P}^{[s]}(\phi)$ denote the union of those
cubes which has at least one face belonging to $s$-th zone. It is
obvious that $\vol(\cZ_{P}^{[s]}(\phi))\leq \vol(\cZ_{P}(\phi))$ for all
$s$.  The main result of this section is the following
\begin{theorem}
 \label{thm:PhysicalBound} If $\phi_{*}\in \Gamma_{n}=\rho^{-1}(r)$ is
 the critical point of $F_{P}$, then
 \[
    \phi^{s}_{*} = \frac{r}{3} \cdot
   \frac{\vol(\cZ_{P}^{[s]}(\phi))}
   {\vol(\cZ_{P}(\phi))}
   ,\qquad (s=1,\dots,n).
 \]
 In particular, the critical point satisfies the inequalities
 \[
   0< \phi_{*}^{i}\leq \frac{r}{3},\quad (i=1,\dots,n).
 \]
 where the equality $\phi_{*}^{i}=\frac{r}{3}$ holds for some $i$ if and
 only if $n=3$ i.e. when the toric diagram $P$ is a triangle.
\end{theorem}

Theorem \ref{thm:PhysicalBound} follows immediately from 
Lemmas \ref{lemma:gradF} and \ref{lemma:vol-zone} below.
\begin{lemma}
\label{lemma:gradF}
 At the critical point $\phi_{*}$, 
 \[
     \frac{\partial \hat{F}_{P}}
    {\partial \phi^{s}}(\phi_{*})=\frac{3}{r}\hat{F}_{P}(\phi_{*})
 \]
 for all $s=1,2,\dots,n$.
\end{lemma}

\Proof
The critical point $\phi_{*}$ is characterized as an extremal point of
the function
\[
  G(\phi)=\hat{F}_{P}(\phi)-\lambda \Bigl(\sum_{i=1}^{n}\phi^{i}-r\Bigr),
\]
where $\lambda$ is a Lagrange multiplier to impose the constraint
$\rho(\phi)=r$. The condition $ dG=0$ leads to
\begin{equation}
 \label{eq:dG=0}
 \frac{\partial \hat{F}_{P}}{\partial \phi^{i}}(\phi_{*}) =\lambda,
 \quad (i=1,\dots,n).
\end{equation}
Since $\hat{F}_{P}$ is homogeneous of degree three, we have
$\sum_{i}\phi^{i} \frac{\partial \hat{F}_{P}(\phi)}{\partial \phi^{i}} =
3\hat{F}_{P}(\phi)$. Multiplying (\ref{eq:dG=0}) by $\phi^{i}$ and
summing over $i$, we have $\lambda=\frac{3}{r}\hat{F}_{P}(\phi_{*})$
which is the desired result. \qed

\begin{lemma}
 \label{lemma:vol-zone} 
 \begin{equation}
  \label{eq:vol-zone} 
  \vol(\cZ_{P}^{[s]}(\phi))=\phi^{s} \frac{\partial \hat{F}_{P}(\phi)}
  {\partial\phi^{s}}, \qquad (s=1,\dots,n).
 \end{equation}
\end{lemma}
\Proof
 Since every term in $\hat{F}_{P}$ defined by (\ref{eq:FP-def}) is at
 most linear in the variable $\phi^{s}$,
 \[
   \vol(\cZ_{P}^{[s]}(\phi))
   =\sum_{\mbox{\scriptsize
   $\matrix{1{\leq} i{<}j{<}k{\leq}n\cr s\in\{i,j,k\}}$}}
   c_{ijk} 
   \phi^{i}\phi^{j}\phi^{k} 
   =\phi^{s} \frac{\partial}{\partial \phi^{s}}
   \sum_{1\leq i<j<k\leq n} c_{ijk}  \phi^{i}\phi^{j}\phi^{k} 
   =\phi^{s} \frac{\partial \hat{F}_{P}(\phi)}{\partial
   \phi^{s}}. 
\]
\qed

It should be noted that summing (\ref{eq:vol-zone}) over $s=1,\dots,n$
and using the homogeneity of $\hat{F}_{P}(\phi)$, we have
$\sum_{s=1}^{n}\vol(\cZ_{P}^{[s]}(\phi))=3\vol(\cZ_{P}(\phi))$.  This
corresponds to the fact that each cube belongs to three distinct zones.

\section{Non-extremal points and monotonicity of the modulus $\mathfrak{M}$}

This section is devoted to the proof of Theorem \ref{thm:main3} (Theorem
\ref{thm:Monotonicity2}).

In the preceding sections, we have assumed that all the vectors
$\v{v}_{1},\v{v}_{2},\dots,\v{v}_{n}$ are extremal points of the toric
diagram $P$. Under this hypothesis, it is shown in Theorem
\ref{thm:main1-again} that the critical point $\phi_{*}$ of $F_{P}$ is
driven away from the boundary of $\Gamma_{n}$. Occasionally we want to
relax this assumption to deal with the geometry such as a suspended
pinch point shown in Figure \ref{fig:SPP}.

\FIGURE{ \epsfig{file=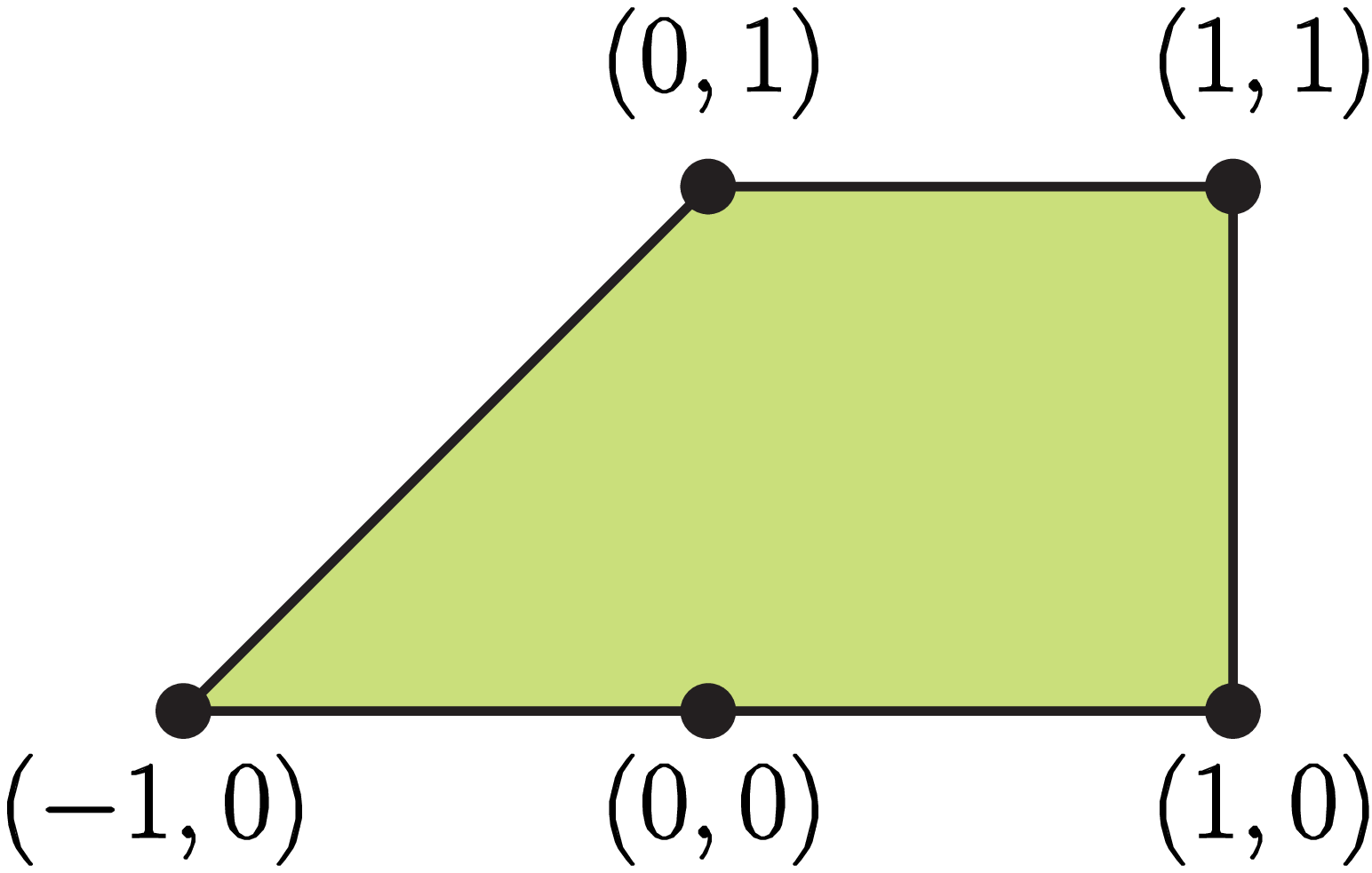,width=.28\textwidth}
 \caption{Suspended pinch point
 \label{fig:SPP}}
}

In this section, we allow $P$ to be simply a set of integral vectors of
a form $(*,*,1)$, not necessarily the vertices (extremal points) of a
convex polygon. Such $P$ will be referred to as a \emph{generalized
toric diagram}; the set of generalized toric diagrams will be denoted by
$\Tg$.  The definitions of the zonotope $Z_{P}(\phi)$, the function
$F_{P}=\vol(\cZ_{P}(\phi))$ and $\mathfrak{M}(P)$ goes through without
any change for any $P\in \Tg$; we agree that the coefficient $c_{ijk}$
is always given by $|\det(\v{v}_{i},\v{v}_{j},\v{v}_{k})|$ even if the
triangle $\triangle(\v{v}_{i}, \v{v}_{j},\v{v}_{k})$ has negative
orientation or contain other $\v{v}_{l}$ inside.

The next Proposition shows that if a non-extremal point $\v{v}_{p}$
exists, the maximization process drives us to the boundary $\phi^{p}=0$. 
Therefore, as far as the maximization of $F_{P}$ is concerned, the
non-extremal points are safely ignored. In physical terms, the
corresponding global symmetry ``decouples'' as a result of
$a$-maximization.
\begin{proposition}
 \label{prop:OnlyVerticesMatter} Let $P=\{\v{v}_{1},\dots,\v{v}_{n}\}\in
 \Tg$ be a generalized toric diagram. Suppose $\v{v}_{p}$ is not an
 extremal point of the convex hull of $P$. Then the volume $F_{P}:
 \Gamma_{n}\to \R$ attains its maximum on the boundary $\partial
 \Gamma_{n}$ corresponding to the hyperplane $\phi^{p}=0$. In other
 words, if we put $P':=P\setminus\{\v{v}_{p}\}=\{\v{v}_{1},\dots,
 \v{v}_{p-1},\v{v}_{p+1},\dots,\v{v}_{n}\}\in \Tg$, then
 \[
    \max_{\phi\in \Gamma_{n}} F_{P}(\phi)
   =\max_{\phi'\in \Gamma_{n-1}} F_{P'}(\phi').
 \]
\end{proposition}
\begin{corollary}
 \label{cor:OnlyVerticesMatter}
 For any $P\in \Tg$, the maximum values of $F_{P}$ is solely
 determined by the convex hull of $P$; if $P$ and $Q$
 are two generalized toric diagrams such that $\conv(P)=\conv(Q)$, then
$  \mathfrak{M}(P)=  \mathfrak{M}(Q)$.
\end{corollary}

\Proof (of Proposition \ref{prop:OnlyVerticesMatter})

\FIGURE{\epsfig{file=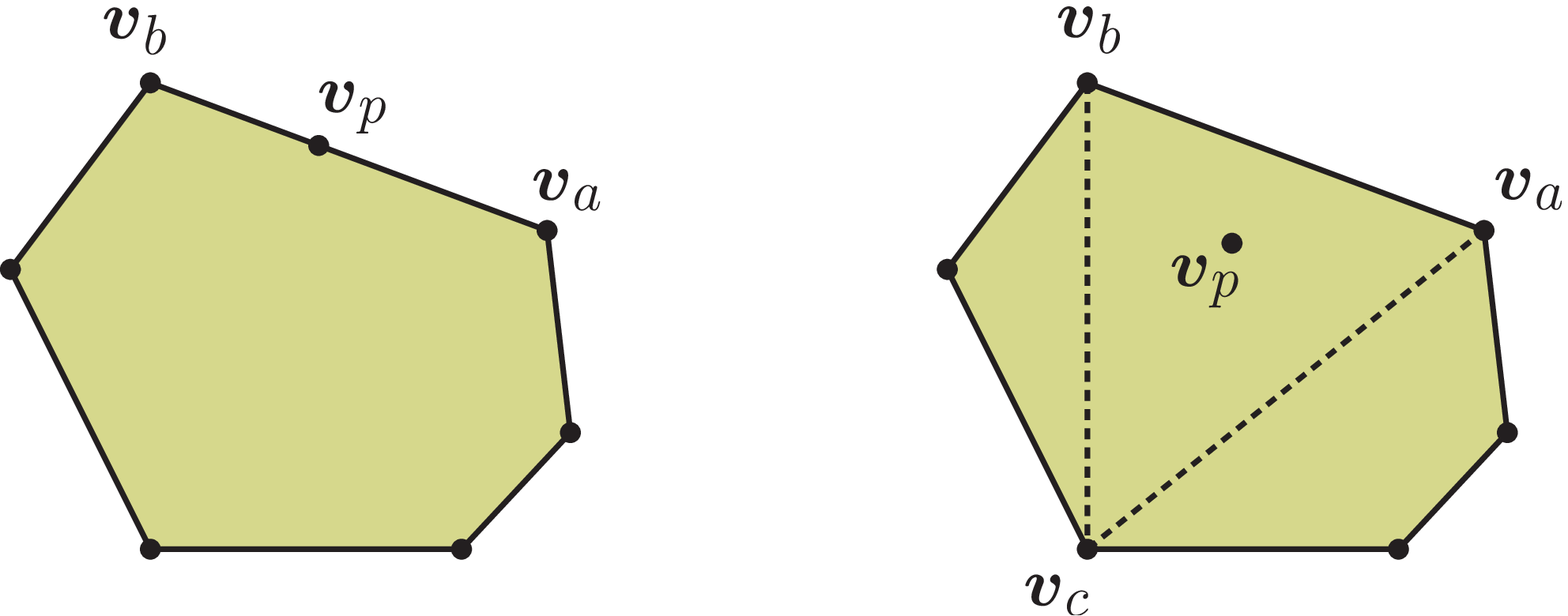,width=.61\textwidth}
  \caption{Left, case (i): $\v{v}_{p}=\alpha \v{v}_{a}+\beta \v{v}_{b}$.
  Right, case (ii) :
  $\v{v}_{p}=\alpha \v{v}_{a}+\beta \v{v}_{b}+\gamma \v{v}_{c}$.
 \label{fig:degen-vector}}
}
 There are two cases to be handled: (i) $\v{v}_{p}\in \partial P$
 and (ii) $\v{v}_{p}\in \relint(P)$ (Figure \ref{fig:degen-vector}).

 \FIGURE[b]{ \epsfig{file=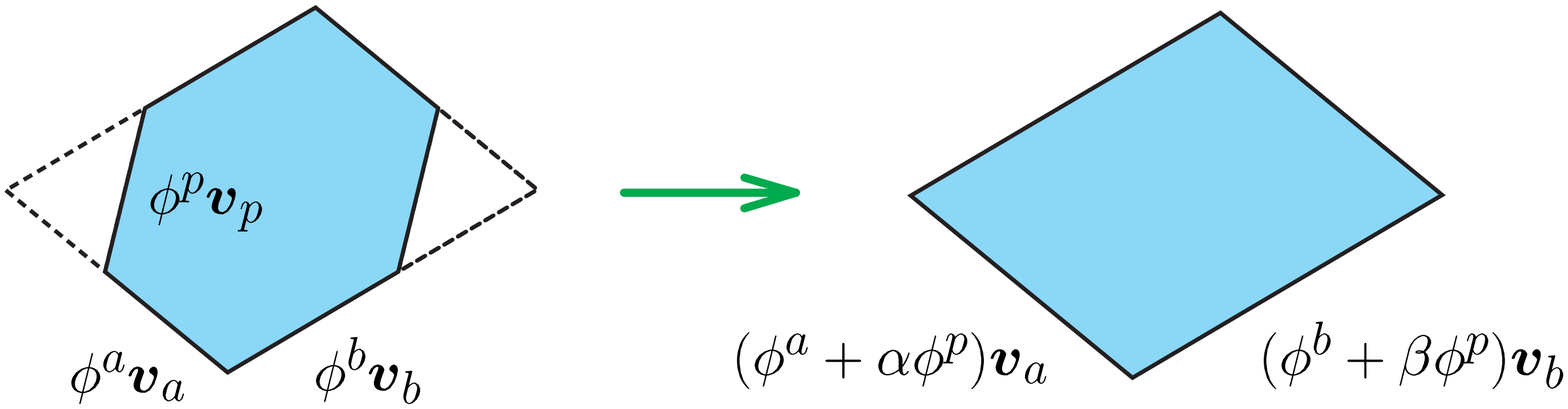,width=0.62\textwidth}
 \caption{Elimination of the non-extremal vector 
 $\v{v}_{p}=\alpha \v{v}_{a}+\beta \v{v}_{b}$. 
 \label{fig:elim-degen-vector}}
 }

 Consider the case (i) first. Let us assume $\v{v}_{p}$ is on a edge
 $\relint(\conv(\v{v}_{a},\v{v}_{b}))$; put $\v{v}_{p}=\alpha
 \v{v}_{a}+\beta \v{v}_{b}$ for some $\alpha,\beta>0$, $\alpha+\beta=1$. 
 Suppose that, contrary to our claim, the maximum of $F_{P}$ is attained
 at $\phi=(\phi^{1},\dots,\phi^{p},\dots,\phi^{n})$ with
 $\phi^{p}>0$. Then we have a following proper inclusion (Figure
 \ref{fig:elim-degen-vector}):
 \begin{eqnarray}
  && \phi^{a}[\v{0},\v{v}_{a}]+
  \phi^{b}[\v{0},\v{v}_{b}]+
  \phi^{p}[\v{0},\v{v}_{p}]
   =
  \phi^{a}[\v{0},\v{v}_{a}]+
  \phi^{b}[\v{0},\v{v}_{b}]+
  \phi^{p}[\v{0},\alpha\v{v}_{a}+\beta\v{v}_{b}]
  \nonumber
   \\
   &&\qquad \qquad 
  \subsetneq
  (\phi^{a}+\alpha \phi^{p})[\v{0}, \v{v}_{a}]
   +  (\phi^{b}+\beta \phi^{p})[\v{0}, \v{v}_{b}].
  \label{eq:ProperInclusion1}
 \end{eqnarray}
 Adding $\sum_{i\neq a,b,p}\phi^{i}[\v{0},\v{v}_{i}]$ to both sides of
 (\ref{eq:ProperInclusion1}), we have $ \cZ_{P}(\phi) \subsetneq
 \cZ_{P}(\phi_{\bullet})$, where the point $\phi_{\bullet}\in
 \partial\Gamma_{n}$ is defined by
 \[
    \phi_{\bullet}^{i}=
   \left\{
   \begin{array}{ll}
    \phi^{a}+\alpha \phi^{p}, \quad& \mbox{if $i=a$},\\
    \phi^{b}+\beta \phi^{p}, & \mbox{if $i=b$},\\
    0, & \mbox{if $i=p$},\\
    \phi^{i}, & \mbox{otherwise}.
   \end{array}
\right.
 \]
 Thus we have $F_{P}(\phi)=\vol(\cZ_{P}(\phi))<
 \vol(\cZ_{P}(\phi_{\bullet})) =F_{P}(\phi_{\bullet})$. More explicitly,
 \[
    F_{P}(\phi_{\bullet})-F_{P}(\phi)
  = \alpha\beta \sum_{i\neq a,b,p}
  |\det(\v{v}_{a},\v{v}_{b},\v{v}_{i})|\phi^{a}\phi^{b}\phi^{i}>0.
 \]
 This contradicts the assumption that $\phi$ is a maximum point. 

 \FIGURE{ \epsfig{file=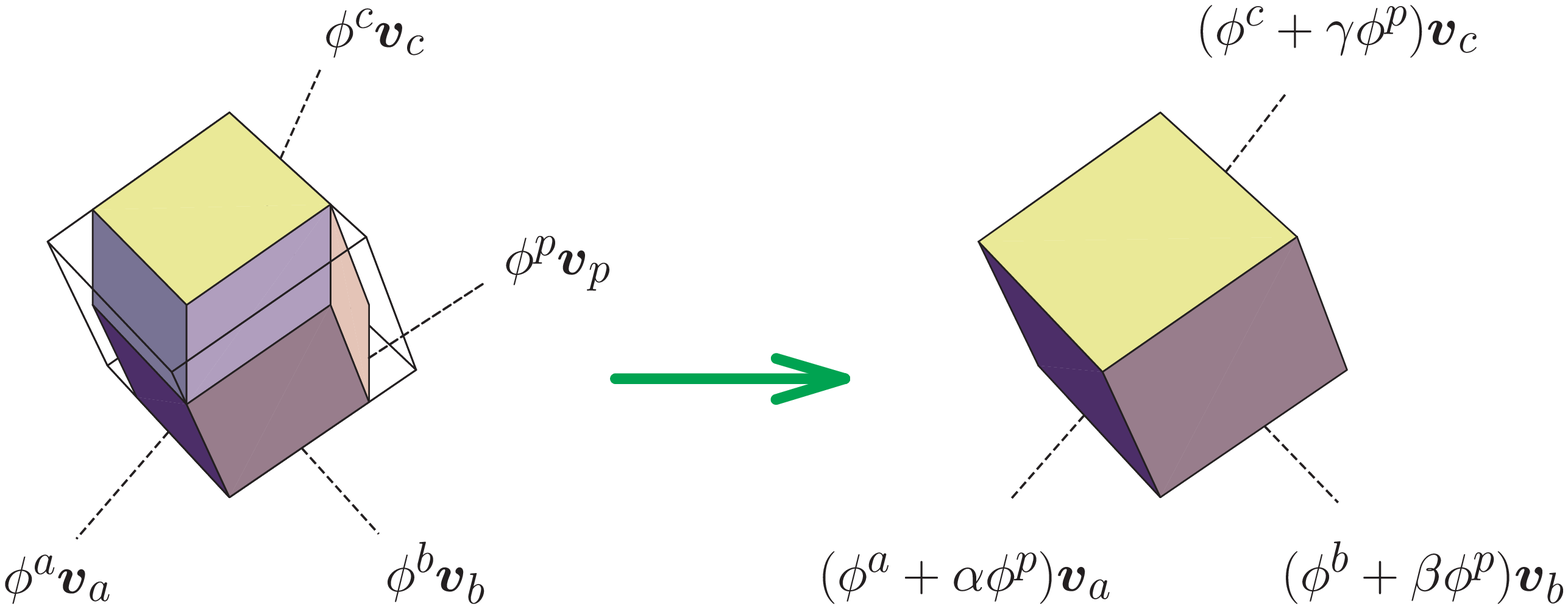,width=0.7\textwidth}
  \caption{Elimination of the non-extremal vector $\v{v}_{p}=\alpha
  \v{v}_{a}+\beta \v{v}_{b}+\gamma \v{v}_{c}$
  \label{fig:elim-degen-vector3}}
  }

 The case (ii) is similar.  There exist three vertices $\v{v}_{a},
 \v{v}_{b},\v{v}_{c}$ ($a,b,c\neq p$) such that $\v{v}_{p}\in \conv(
 \v{v}_{a}, \v{v}_{b},\v{v}_{c})$. Let $\alpha,\beta,\gamma$ be three
 nonnegative numbers such that $\alpha+\beta+\gamma=1$ and
 $\v{v}_{p}=\alpha \v{v}_{a}+\beta \v{v}_{b}+\gamma \v{v}_{c}$. To
 obtain a contradiction, suppose that $F_{P}$ takes its maximum at
 $\phi=(\phi^{1},\dots,\phi^{p},\dots,\phi^{n})\in \Gamma_{n}$ with
 $\phi^{p}>0$. We have a following proper inclusion (Figure
 \ref{fig:elim-degen-vector3})
   \begin{eqnarray}
  &&\phi^{a}[\v{0},\v{v}_{a}]+
  \phi^{b}[\v{0},\v{v}_{b}]+
  \phi^{c}[\v{0},\v{v}_{c}]+
  \phi^{p}[\v{0},\v{v}_{p}]
  \nonumber
  \\
   && =
  \phi^{a}[\v{0},\v{v}_{a}]+
  \phi^{b}[\v{0},\v{v}_{b}]+
  \phi^{c}[\v{0},\v{v}_{c}]+
  \phi^{p}[\v{0},\alpha\v{v}_{a}+\beta\v{v}_{b}+\phi^{c}\v{v}_{c}]
  \nonumber
   \\
   &&
  \subsetneq
  (\phi^{a}+\alpha \phi^{p})[\v{0}, \v{v}_{a}]
   +  (\phi^{b}+\beta \phi^{p})[\v{0}, \v{v}_{b}]
   +(\phi^{a}+\gamma\phi^{c})[\v{0},\phi^{c}\v{v}_{c}].
   \label{eq:ProperInclusion2}
   \end{eqnarray}
 Adding $\sum_{i\neq a,b,c,p}\phi^{i}[\v{0},\v{v}_{i}]$ to both sides of
 (\ref{eq:ProperInclusion2}), we have $\cZ_{P}(\phi) \subsetneq
 \cZ_{P}(\phi_{\bullet})$ where the point $\phi_{\bullet}\in
 \partial\Gamma_{n}$ is defined by
 \[
    \phi_{\bullet}^{i}=
\left\{
   \begin{array}{ll}
    \phi^{a}+\alpha \phi^{p}, \quad&\mbox{if $i=a$},\\
    \phi^{b}+\beta \phi^{p}, & \mbox{if $i=b$},\\
    \phi^{c}+\gamma \phi^{p}, & \mbox{if $i=c$},\\
    0, & \mbox{if $i=p$},\\
    \phi^{i}, & \mbox{otherwise}.
   \end{array}
\right.
 \]
 Thus we have $F_{P}(\phi)=\vol(\cZ_{P}(\phi))<
 \vol(\cZ_{P}(\phi_{\bullet}))=F_{P}(\phi_{\bullet})$.  More explicitly,
 the volume increases by
 \begin{eqnarray*}
   F_{P}(\phi_{\bullet})-F_{P}(\phi)
   &=& 
   \alpha\beta\gamma \;
   |\det(\v{v}_{a},\v{v}_{b},\v{v}_{c})|\phi^{a}\phi^{b}\phi^{c}
   +
   \alpha\beta \sum_{i\neq a,b,p}
   |\det(\v{v}_{a},\v{v}_{b},\v{v}_{i})|\phi^{a}\phi^{b}\phi^{i}
  \\
  && 
   +
   \beta\gamma \sum_{i\neq b,c,p}
   |\det(\v{v}_{b},\v{v}_{c},\v{v}_{i})|\phi^{b}\phi^{c}\phi^{i}
   +
   \gamma\alpha \sum_{i\neq c,a,p}
   |\det(\v{v}_{c},\v{v}_{a},\v{v}_{i})|\phi^{c}\phi^{a}\phi^{i}
   > 0.
 \end{eqnarray*}
 This is a contradiction.
\qed

We investigate a few more ways of changing toric diagrams.
\begin{proposition}
 \label{prop:Monotonicity-2} Suppose a toric diagram $P$ is obtained
 from a toric diagram $Q$ by elongating one or two edges of $Q$ as in
 Figure \ref{fig:elongatingP}. Then, $\mathfrak{M}(Q)<\mathfrak{M}(P).$
\end{proposition}
\FIGURE{ \epsfig{file=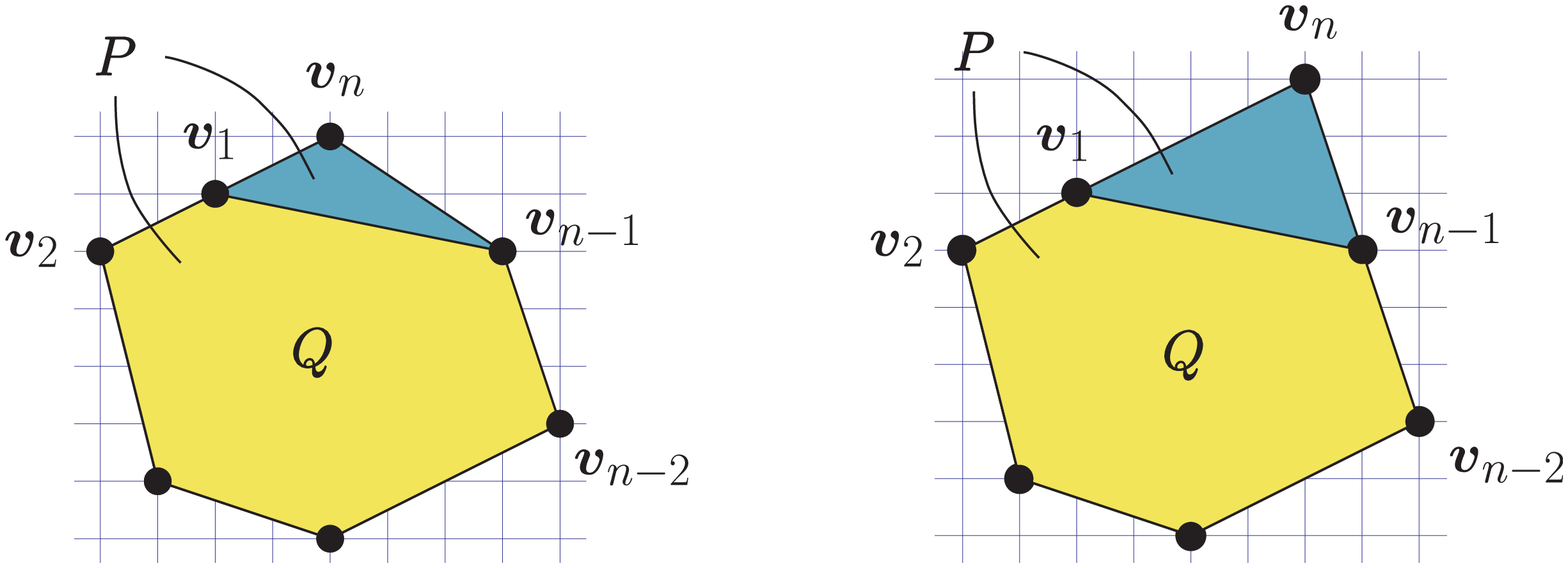,width=.67\textwidth}
 \caption{Elongating one or two edges of a toric diagram
 \label{fig:elongatingP}}
}
\Proof
 Consider the toric diagram $P$ in Figure \ref{fig:elongatingP} on the
 left. $\v{v}_{1}$ is non-extremal point of $P$, thus
\begin{eqnarray*}
\mathfrak{M}(P)&
   =&\mathfrak{M}(\{\v{v}_{2},\dots,\v{v}_{n-1},\v{v}_{n}\})\\
   &=&\mathfrak{M}(\{\v{v}_{1},\v{v}_{2},\dots,\v{v}_{n-1},\v{v}_{n}\})
   \qquad \mbox{by Corollary \ref{cor:OnlyVerticesMatter} }\\
   &>&\mathfrak{M}(\{\v{v}_{1},\v{v}_{2},\dots,\v{v}_{n-1}\})
   \qquad~~~~~ \mbox{by Corollary \ref{cor:Monotonicity} }\\
   &=&\mathfrak{M}(Q).
\end{eqnarray*}   
 The toric diagrams on the right of Figure \ref{fig:elongatingP} can
 be handled in much the same way. 
\qed

\begin{proposition}
 \label{prop:PushOut} Suppose a toric diagram $P$ is obtained from a
 toric diagram $Q$ by pushing out one vertex of $Q$ as depicted in
 Figure \ref{fig:PushOut} on the left. Then,
 $\mathfrak{M}(Q)<\mathfrak{M}(P)$.
\end{proposition}
\FIGURE{ \epsfig{file=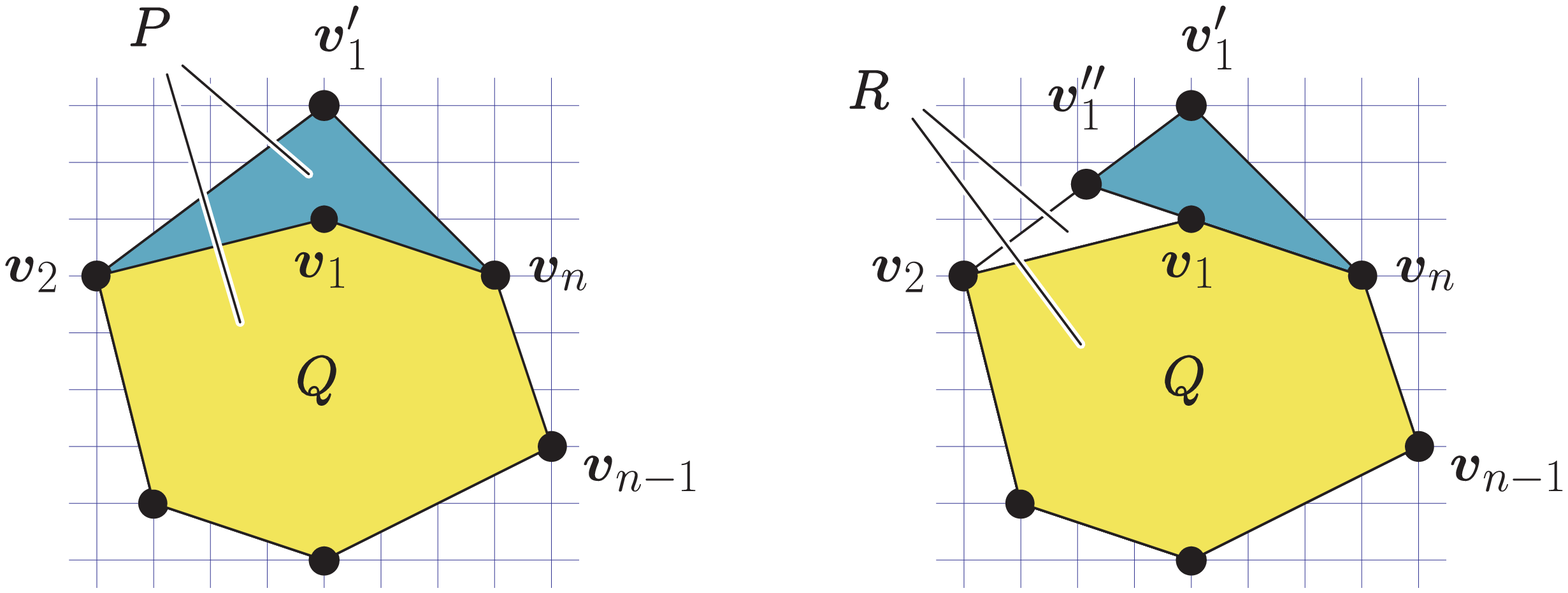,width=.67\textwidth}
 \caption{Pushing out a vertex of a toric diagram
 \label{fig:PushOut}}
}
\Proof
 Label the vertices of $P$ and $Q$ as in Figure
 \ref{fig:PushOut}. Construct another polygon\footnote{The coordinates
 of the vertex $\v{v}''_{1}$ are rational numbers in general; the
 polygon $R$ is not a toric diagram in the strict sense. The conclusion
 is true, however, because all the results used here are proved with no
 assumption on the integrality of the vertices.}
 $R=\{\v{v}''_{1},\v{v}_{2},\dots$ $,\v{v}_{n-1},\v{v}_{n}\}$ by
 extending the edge $(\v{v}_{n},\v{v}_{1})$ into the direction of
 $\v{v}_{1}$ until it touches the edge $(\v{v}'_{1},\v{v}_{2})$ at
 $\v{v}''_{1}$ (Figure \ref{fig:PushOut}, right). Clearly $Q\subsetneq
 R\subsetneq P$. Applying Proposition \ref{prop:Monotonicity-2} twice,
 we have $\mathfrak{M}(Q)<\mathfrak{M}(R)<\mathfrak{M}(P)$. 
\qed

Now suppose $P$ and $P'$ are two toric diagrams satisfying $P\subset
P'$. It is clear that there is a sequence of toric diagrams (or rational
polygons)
\begin{equation}
 P=Q_{0}\subset Q_{1}\subset Q_{2}\subset
  \dots
\subset Q_{k-1}\subset Q_{k}=P'
\end{equation}
such that $Q_{i}$ ($1\leq i\leq k$) is obtained from $Q_{i-1}$, either
\begin{itemize}
 \itemsep=0mm
 \item by adding an extra vertex to $Q_{i-1}$ 
       (Proposition \ref{prop:Monotonicity}),
 \item by elongating one or two edges
       of $Q_{i-1}$ (Proposition \ref{prop:Monotonicity-2}), or
 \item by pushing out one vertex of $Q_{i-1}$ 
	    (Proposition \ref{prop:PushOut}).
\end{itemize}
In either case, we know that
$\mathfrak{M}(Q_{i-1})<\mathfrak{M}(Q_{i})$.  Consequently, we have
demonstrated the following monotone property of the maximum value of
$F_{P}$, or modulus $\mathfrak{M}(P)$, with respect to the change of the
toric diagrams:
\begin{theorem}
 \label{thm:Monotonicity2} Let $P$ and $P'$ be two toric diagrams
 satisfying $[P]\preceq [P']$, where $\preceq$ is the partial order defined in
 (\ref{eq:PartialOrder}). Then $\mathfrak{M}(P)\leq
 \mathfrak{M}(P')$. The equality holds if and only if $P\simeq P'$,
 i.e. equal up to integral affine transformations on $\Z^{2}$.
\end{theorem}

\section{Relation to volume minimization}
\label{sec:vol-min}

In the preceding sections we have concerned ourselves with the
extremization of homogeneous polynomial $\hat{F}_{P}:\R_{\geq 0}^{n}\to
\R$ under the constraint
\begin{equation}
\label{eq:phisum=r}
 \rho(\phi)= \phi^{1}+\dots+\phi^{n}=r.
\end{equation}
Now consider a following variant of the extremization problem.

\DOUBLEFIGURE[b]{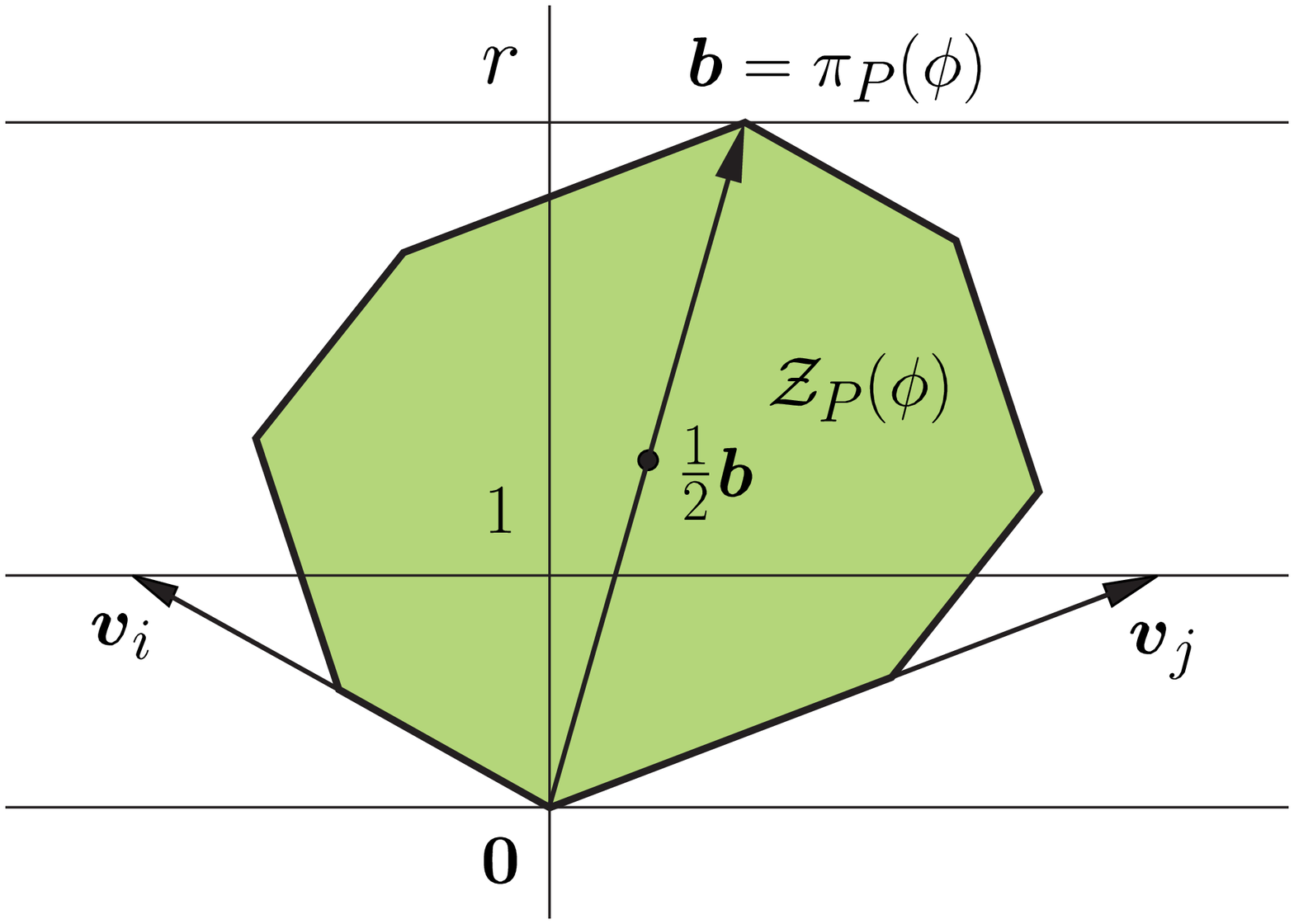,width=.40\textwidth}{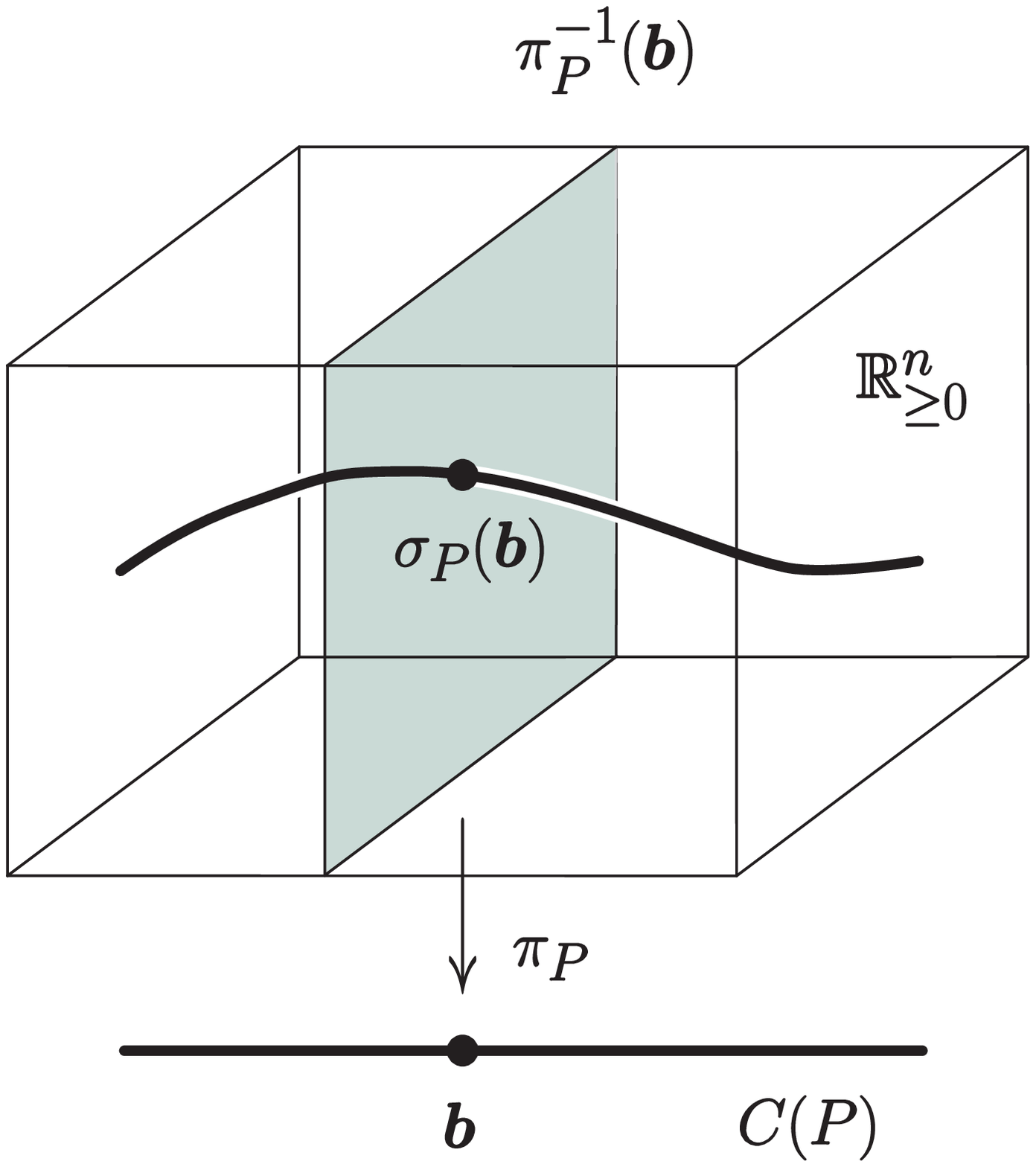,width=.34\textwidth}
{Zonotope and Reeb vector $\v{b}$\label{fig:reeb}}{Fibration $\pi_{P}:\R_{\geq 0}^{n}\to C(P)$  and fiberwise
 critical point $\sigma_{P}(\v{b})$.\label{fig:fibration}}

As before, let $P\in \T$ be a toric diagram and
$\v{v}_{1},\dots,\v{v}_{n}$ be its vertices in counter\-clock\-wise
order. Define a map $ \pi_{P}:\R_{\geq 0}^{n} \to\R^{3} $ by
\[
  \pi_{P}(\phi) = \sum_{i=1}^{n}\phi^{i}\v{v}_{i}.
\]
The image of $\pi_{P}$ is the cone $C(P)$ over the toric diagram $P$
(Figure \ref{fig:toric-diag-cone}). In terms of the zonotope
$\cZ_{P}(\phi)$, $\v{b}=\pi_{P}(\phi)$ is the location of the vertex
opposite to the origin $\v{0}$, and $\frac{1}{2}\v{b}$ is the center
(Figure \ref{fig:reeb}). As we will see, $\v{b}\in C(P)$ can be
identified with the Reeb vector of a Sasaki-Einstein manifold.

Due to the fact that each $\v{v}_{i}$ is of the form $(*,*,1)$, the
constraint (\ref{eq:phisum=r}) factors through the projection $\pi_{P}$:
\[
   \arraycolsep=2pt
 \begin{array}{ccccccc}
  \rho & : & \R_{\geq 0}^{n} & \stackrel{\pi_{P}}{\longrightarrow} & C(P) &
 \stackrel{h}{\longrightarrow} & \R \\
  & & \cup & & \cup & & \cup \\
  & & \Gamma_{n}&\stackrel{\pi_{P}}{\longrightarrow} 
   & rP & \stackrel{h}{\longrightarrow}& \{r\}\\
  \end{array}
\]
where $h(x,y,z)=z$. The image of $\Gamma_{n}$ under $\pi_{P}$ is
$rP=h^{-1}(r)$, namely the horizontal slice of the cone $C(P)$ at height
$r$.

For a generic $\v{b}\in C(P)$, the fiber $\pi_{P}^{-1}(\v{b})$ of the
projection $\pi_{P}:\R_{\geq 0}^{n} \to C(P)$ is an
$(n{-}3)$-dimensional convex polytope.  This fibration structure allows
us to extremize $F_{P}:\Gamma_{n}\to \R$ in two steps: first in the
fiber direction, and then in the base direction (Figure
\ref{fig:fibration}):
\[
  \max_{\phi\in \Gamma_{n}} \hat{F}_{P}(\phi) = \max_{\strut\v{b}\in rP}
  \biggl\{
  \max_{\strut\phi\in \pi_{P}^{-1}(\v{b})} \hat{F}_{P}(\phi)
  \biggr\}.
\]

The following theorem shows the maximization along the fiber direction
is rather straightforward and admits an explicit solution.

\begin{theorem}
 \label{thm:FiberDirectionQuadratic} 
 ~
 Let $r>0$ and $\v{b}$ be a point in $rP\subset C(P)$.
\begin{itemize}
 \item [(i)] The restriction
       $\hat{F}_{P}|_{\pi_{P}^{-1}(\v{b})}$ of $\hat{F}_{P}$ along the
       fiber $\pi_{P}^{-1}(\v{b})$ is a quadratic polynomial.

 \item [(ii)] The quadratic polynomial
       $\hat{F}_{P}|_{\pi_{P}^{-1}(\v{b})}$ has a unique critical point
       $\sigma_{P}(\v{b})$ and it is also a maximum.  The point
       $\sigma_{P}(\v{b})$ is determined as follows: Let $\ell_{P} :
       \R^{3}\to \R^{n}$ be a vector-valued rational function defined by
       \[
		 \ell_{P}^{i}(\v{b}):=
	 \frac{\vev{\v{v}_{i-1},\v{v}_{i},\v{v}_{i+1}}}
	 {\vev{\v{b},\v{v}_{i-1},\v{v}_{i}}
	 \vev{\v{b},\v{v}_{i},\v{v}_{i+1}}},\qquad
	 (\v{b}\in \R^{3},~~i=1,\dots,n).
       \]
       Then 
       \begin{equation}
	\label{eq:FiberCritPosition}
	\sigma_{P}^{i}(\v{b}) = \frac{r}{V_{P}(\v{b})} \ell_{P}^{i}(\v{b})
	,\qquad  (i=1,\dots,n)
       \end{equation}
       where
       \[
		 V_{P}(\v{b}):=\sum_{i=1}^{n}\ell_{P}^{i}(\v{b}).
       \]
       The critical value is given by
       \[
       \max_{\phi\in \pi_{P}^{-1}(\v{b})} \hat{F}_{P}(\phi) =
	\hat{F}_{P}(\sigma_{P}(\v{b}))= \frac{r}{V_{P}(\v{b})}.
       \]
 \end{itemize}
\end{theorem}

Before we turn to the proof of Theorem
\ref{thm:FiberDirectionQuadratic}, it is useful to take a small detour
into the AdS/CFT correspondence and the volume minimization.

The AdS/CFT-correspondence
\cite{Maldacena:1997re,Gubser:1998bc,Witten:1998qj,Aharony:1999ti} has
brought a novel insight into the relation between $N{=}1$ SCFT and
Sasaki-Einstein manifolds. The world-volume theory on the D3 branes
living at conical Calabi-Yau singularities is dual to a type IIB
background of the form $AdS_5\times Y$, where $Y$ is the five
dimensional horizon manifold
\cite{Klebanov:1998hh,Acharya:1998db,Morrison:1998cs}. Supersymmetry
requires that $Y$ is a Sasaki-Einstein manifold and the cone $X=C(Y)$
over the base $Y$ is a Calabi-Yau threefold with Gorenstein singularity.

If $Y$ is toric, i.e. its isometry group contains at least three torus,
then $X$ is a toric Calabi-Yau singularity.  It is known \cite{MR567836}
that every toric Gorenstein Calabi-Yau singularity is obtained as a
toric variety $X_{P}$ associated with a fan $C(P)$, the cone over a
toric diagram $P$.  The toric variety $X_{P}$ is equipped with a moment
map $\mu:X_{P}\to \R^{3}$ associated with the $T^{3}\subset
(\C^{*})^{3}$ action. The image of $\mu$ is the dual cone $C(P)^{\vee}$,
and the generic fiber of $\mu$ is $T^{3}$. The corresponding
Sasaki-Einstein manifold $Y_{P}$ has a canonically defined constant norm
Killing vector field, called Reeb vector field; it is identified with a
vector $\v{b}$ in $C(P)$ via moment map $\mu$. Each vertex $\v{v}_{i}$
of $P$ determines a toric divisor $D_{i}$ in $X_{P}$, which is a cone
over a certain calibrated three dimensional submanifold $\Sigma_{i}$ in
$Y_{P}$.

The Calabi-Yau cone $X_{P}$ can be also constructed as a symplectic
quotient \cite{MR1293656,MR2039164,MR984900}
(up to some finite abelian group)
\begin{equation}
 \label{eq:QuotientConstruction}
 X_{P}\simeq \C^{n}//(\C^{*})^{n-3}.
\end{equation}
The standard $(\C^{*})^{n}$ action on $\C^{n}$ can be decomposed into
$(\C^{*})^{n-3}$ and $(\C^{*})^{3}$ corresponding to the exact sequence
\[
  0\longrightarrow \R^{n-3}\longrightarrow 
  \R^{n} \stackrel{\pi_{P}}{\longrightarrow} \R^{3}
  \longrightarrow 0.
\]
The $(\C^{*})^{n-3}$ action, defining the quotient action in
(\ref{eq:QuotientConstruction}), is called baryonic symmetries in
physics literature; $(\C^{*})^{3}$ acting nontrivially on $X_{P}$ is
referred to as flavor symmetries. The flavor (resp. baryonic) symmetry
corresponds to the base (resp. fiber) direction of the fibration
$\pi_{P}:\R_{\geq 0}^{n}\to C(P)$.

According to the prediction of AdS/CFT correspondence, the central
charge $a$ of the SCFT and the volume of the internal manifold are
related as \cite{Henningson:1998gx,Gubser:1998vd}
\[
  a=\frac{N^{2}\pi^{3}}{4\vol(Y)},
\]
while the exact $R$-charges of chiral fields are proportional to the volumes
of three cycles $\Sigma_{i}\subset Y_{P}$ 
 \cite{Gubser:1998fp}
\[
  R_{i}=\frac{\pi}{3} \frac{\vol(\Sigma_{i})}{\vol(Y)}.
\]

It is usually quite difficult to obtain Einstein metrics explicitly;
it thus appears impossible to compute these volumes. Remarkably,
Martelli, Sparks and Yau  \cite{Martelli:2005tp,Martelli:2006yb} proved
that the volumes of $Y_{P}$ and $\Sigma_{i}$'s can be computed without
actually knowing the metric, provided the Reeb vector $\v{b}\in 3P$ is
known for the Calabi-Yau cone\footnote{The number $3$ of $\v{b}\in 3P$
is due to the fact that $\dim_{\C} X_{P}=3$.}:
\begin{eqnarray*}
  \vol(\Sigma_{i})&=&{2\pi^{2}}\ell_{P}^{i}(\v{b}),
  \\
  \vol(Y_{P})&=&\frac{\pi}{6}\sum_{i=1}^{n} \vol(\Sigma_{i})
  =\frac{\pi^{3}}{3}V_{P}(\v{b}).
\end{eqnarray*}
Here, the functions $\ell_{P}^{i}(\v{b})$ and $V_{P}(\v{b})$ of $\v{b}$
are nothing but those defined in Theorem
\ref{thm:FiberDirectionQuadratic}.  Even more importantly, these authors
showed that the correct Reeb vector is characterized as the vector
$\v{b}\in 3P$ which minimize the ``trial volume function''
$V_{P}(\v{b})$. This is a beautiful geometrical counterpart of the
$a$-maximization.

It should be clear now how the volume minimization and
``$a$-maximization in two steps'' are related. Theorem
\ref{thm:FiberDirectionQuadratic} can be summarized as the following
\begin{theorem}
 \label{thm:BZ}

 (i) For $\v{b}=(*,*,r)\in \relint(rP)$, the trial volume function
 $V_{P}(\v{b})$ is inversely proportional to the maximum of the
 $a$-function in the fiber $\pi_{P}^{-1}(\v{b})$:
 \[
    \max_{\phi\in \pi_{P}^{-1}(\v{b})}
  F_{P}(\phi)=\frac{r}{V_{P}(\v{b})}.
 \]

 (ii) The $a$-maximization and the volume minimization are 
 equivalent in the sense that 
 \[
    \max_{\phi\in \rho^{-1}(r)} F_{P}(\phi) = \frac{r}{\displaystyle
   \min_{\v{b}\in \relint (rP)} V_{P}(\v{b})}.
 \]
\end{theorem}
Theorem \ref{thm:BZ} is essentially due to Butti-Zaffaroni
\cite{Butti:2005vn}. Our approach has an advantage that the existence of
the maximum in each fiber is almost immediate from the concavity.
\footnote{In \cite{Butti:2005vn} it was shown that the extremal point of
$\hat{F}_{P}(\phi)$ corresponds to the minimum of $V_{P}(\v{b})$; but
whether the extremal point is a local maximal or not remained open in
their paper.}

Let us now turn to the proof of Theorem \ref{thm:FiberDirectionQuadratic}.  
We first prove the following 
\begin{lemma}
 \label{prop:F-s-factored}
 \[
    \hat{F}_{P}(\phi) = \biggl(\sum_{i=1}^{n}\phi^{i}\v{v}_{i}\biggr) \cdot
   \biggl(\sum_{1\leq j<k\leq n}^{n}\phi^{j}\phi^{k}\v{v}_{j}\times
   \v{v}_{k}\biggr).
 \]
\end{lemma}
\Proof
 The right hand side can be rewritten as follows:
 \begin{eqnarray*}
   \lefteqn{ \sum_{i=1}^{n} \sum_{1\leq j<k\leq n}^{n}
   \phi^{i}\phi^{j}\phi^{k}\vev{\v{v}_{i}, \v{v}_{j}, \v{v}_{k}}}
   \\
   & = &\biggl(\sum_{1\leq i<j<k\leq n}^{n}
   + \sum_{1\leq j<i<k\leq n}^{n} + \sum_{1\leq j<k<i\leq n}^{n}\biggr)
   \phi^{i}\phi^{j}\phi^{k}\vev{\v{v}_{i}, \v{v}_{j}, \v{v}_{k}}
   \\
   & =  &  \sum_{1\leq i<j<k\leq n}^{n}
   \phi^{i}\phi^{j}\phi^{k}
   \bigl(
   \vev{\v{v}_{i}, \v{v}_{j}, \v{v}_{k}}+
   \vev{\v{v}_{j}, \v{v}_{i}, \v{v}_{k}}+
   \vev{\v{v}_{k}, \v{v}_{i}, \v{v}_{j}}
   \bigr)
   \\
   &= &
   \sum_{1\leq i<j<k\leq n}^{n}
   \phi^{i}\phi^{j}\phi^{k}
   \bigl(
   \vev{\v{v}_{i}, \v{v}_{j}, \v{v}_{k}}
   -\vev{\v{v}_{i}, \v{v}_{j}, \v{v}_{k}}
   +\vev{\v{v}_{i}, \v{v}_{j}, \v{v}_{k}}
   \bigr)
   \\
   &
   =&
   \sum_{1\leq i<j<k\leq n}^{n}
   \phi^{i}\phi^{j}\phi^{k}
   \vev{\v{v}_{i}, \v{v}_{j}, \v{v}_{k}}
  = \hat{F}_{P}(\phi).
 \end{eqnarray*}
\qed

The first claim of Theorem \ref{thm:FiberDirectionQuadratic} follows
immediately from Lemma \ref{prop:F-s-factored}; the restriction of
$\hat{F}_{P}$ to the fiber $\pi_{P}^{-1}(\v{b})$ equals a quadratic
polynomial
\[
   Q(\phi):=  \sum_{1\leq j<k\leq n}^{n}
   \vev{\v{b},\v{v}_{j}, \v{v}_{k}} \phi^{j}\phi^{k},
   \qquad (\phi \in \pi_{P}^{-1}(\v{b})).
\]
To compute the critical point of $Q$, introduce an $n\times n$ symmetric matrix
$A=(A_{ij})$ defined by
\[
    A_{ij}=A_{ji}=\frac{1}{2}\vev{\v{b},\v{v}_{i}, \v{v}_{j}},
   \qquad (1\leq i\leq j\leq n)
\]
so that $Q(\phi)=\sum_{i,j=1}^{n}A_{ij}\phi^{i}\phi^{j}$.

The extremization of $Q:\pi_{P}^{-1}(\v{b})\to \R$ is equivalent to that
of a function $\tilde{Q}$ defined by
\[
   \tilde{Q}=\sum_{i,j=1}^{n}A_{ij}\phi^{i}\phi^{j} -\v{\lambda}\cdot
   \bigl(\sum_{i=1}^{n}\phi^{i}\v{v}_{i}-\v{b}\bigr)
\]
where $\v{\lambda}\in \R^{3}$ is the Lagrange multiplier imposing the
constraint $\pi_{P}(\phi)=\v{b}$. The equation $d\tilde{Q}=0$ gives
\begin{equation}
  \label{eq:CritOfFH}
  2 \sum_{j=1}^{n}A_{ij}\sigma_{P}(\v{b})^{j}
  -\v{\lambda}\cdot \v{v}_{i}=0,\qquad (i=1,\dots,n)
\end{equation}
or equivalently,
\begin{equation}
  \label{eq:sigma-A-lambda}
  \sigma_{P}(\v{b})^{i}=\frac{1}{2}\sum_{j=1}^{n}
   (A^{-1})^{ij} \v{\lambda}\cdot \v{v}_{j}.\qquad (i=1,\dots,n)
\end{equation}
In Appendix, the inverse matrix $A^{-1}$, which exists if $\v{b}\in
\relint(C(P))$, is explicitly calculated. Applying Proposition
\ref{prop:Ainverse} (i) to (\ref{eq:sigma-A-lambda}), we have
\begin{eqnarray}
   \sigma_{P}(\v{b})^{i} & = &
   \frac{1}{2}\biggl[
   (A^{-1})^{i\,i{-}1} \v{\lambda}\cdot \v{v}_{i-1}
   +(A^{-1})^{i\,i} \v{\lambda}\cdot \v{v}_{i}
   +(A^{-1})^{i\,i{+}1} \v{\lambda}\cdot \v{v}_{i+1}
   \biggr]
   \nonumber
   \\
   &= &\frac{1}{2}
     \frac{\vev{\v{v}_{i-1},\v{v}_{i},\v{v}_{i+1}}}
     {\vev{\v{b},\v{v}_{i-1},\v{v}_{i}}\vev{\v{b},\v{v}_{i},\v{v}_{i+1}}} 
   \v{\lambda}\cdot \v{b}
   \nonumber
   \\
   &=& \frac{1}{2}(\v{\lambda}\cdot \v{b}) \;\ell_{P}^{i}(\v{b}),
   \qquad (i=1,\dots,n).
 \label{eq:FiberCrit}
\end{eqnarray}
Summing over $i$ and using the constraint
$r=\sum_{i=1}^{n}\sigma_{P}(\v{b})^{i}$, one has
\begin{equation}
  \label{eq:lambda-s} \frac{1}{2}(\v{\lambda}\cdot \v{b}) = \frac{r}{
   \sum_{i=1}^{n} \ell_{P}^{i}(\v{b})}
   =\frac{r}{V_{P}(\v{b})}.
\end{equation}
The formula (\ref{eq:FiberCritPosition}) follows from
(\ref{eq:FiberCrit}) and (\ref{eq:lambda-s}). The critical value is then
given by, from (\ref{eq:CritOfFH}) and (\ref{eq:lambda-s}),
\begin{eqnarray*}
 Q(\sigma_{P}(\v{b}))&=& \sum_{i,j=1}^{n}
  \sigma_{P}(\v{b})^{i}(A_{ij}\sigma_{P}(\v{b})^{j})
  = \frac{1}{2}\sum_{i=1}^{n}\sigma_{P}(\v{b})^{i}(\v{\lambda}\cdot \v{v}_{i})
 \\
 & =&  \frac{1}{2} \v{\lambda}\cdot \pi_{P}(\sigma_{P}(\v{b}))
 =  \frac{1}{2} \v{\lambda}\cdot \v{b}
  = \frac{r}{V_{P}(\v{b})}.
\end{eqnarray*}
The critical point $\sigma(\v{b})$ of $Q$ is easily seen to be a maximum
along the similar lines as the proof of Proposition
\ref{prop:CriticalIsUnique}.  This completes the proof of Theorem
\ref{thm:FiberDirectionQuadratic}.

\section{Summary and outlook}

In this paper, we proved that for any toric diagram $P$, the
$a$-maximization always leads to the unique solution, which satisfies a
universal upper bound. A combinatorial analogue of $a$-theorem is also
established: the $a$-function always decreases when a toric diagram gets
smaller. The relation between $a$-maximization and volume minimization
is also discussed.

A tacit assumption in this paper is that a quiver gauge theory is
uniquely determined by a toric diagram. The formulae (\ref{eq:a-def})
and (\ref{eq:Cijk-def}) are associated with toric diagrams so naturally
that there seems to be no other choice.  However, the brane-tiling
technique \cite{Hanany:2005ve,Franco:2005rj} allows us to produce many
examples of gauge theories whose matter content is different from what
we studied in this paper; they are regarded as realizing different
``phases'' of the same SCFT.  The gauge theory studied in this paper is
called ``minimal'' in \cite{Butti:2005vn}; conjecturally the number of
chiral fields given in (\ref{eq:Ns-from-P}) is smaller than that of any
other possible phases. It is interesting to study $a$-maximization of
those non-minimal phases and compare with the minimal ones.

Since the polynomial $F_{P}$ is defined over the integers,
the maximum value of $F_{P}$ or $\mathfrak{M}(P)$ is always an algebraic
number. Does this critical value characterize SCFT uniquely? In two
dimensions, there are examples of non-isomorphic CFTs with equal
Virasoro central charges. The situation is not clear for higher
dimensions. As we have seen, the $a$-maximization defines a natural map
\[
  \mathfrak{M} \;:\; \T/ \simeq \quad 
  \longrightarrow \quad 
  \overline{\Q} \cap \R_{\geq 0}.
\]
Theorem \ref{thm:Monotonicity2} asserts that $\mathfrak{M}$ is strictly
decreasing along any descending chain (totally ordered subset) of
$\T/\simeq$. Although $\mathrm{Area}(P)$ shares this property, they just
encode the number of gauge fields; there are many toric diagrams of the
same area but different modulus. It seems that modulus
$\mathfrak{M}$ is far more sensitive to the shape of toric diagrams than
the area. We conjecture that the map $\mathfrak{M}$ is injective.

Probably the most important question is why zonotopes come into play in
$a$-max\-i\-mi\-za\-tion. We believe that a deeper understanding of this
will shed new light on the AdS/CFT correspondence.

\acknowledgments

We are grateful to H. Fuji, Y. Imamura, Y. Nakayama, T. Sakai,
Y. Tachikawa, T. Takayanagi and M. Yamazaki for many interesting
discussions and useful comments. This work is supported by Grants-in-Aid
for Scientific Research and the Japan Society for Promotion of Science
(JSPS).

\appendix
\section{Appendix: A symmetric matrix and its inverse}

\begin{proposition}
 \label{prop:Ainverse} Let $\v{b},\v{v}_{1},\dots,\v{v}_{n}$ be vectors
 in $\R^{3}$ and $A=(A_{ij})$ be an $n\times n$ symmetric matrix given
 by
 \[
    A_{ij}=A_{ji}=\frac{1}{2}\vev{\v{b},\v{v}_{i}, \v{v}_{j}},
   \qquad (1\leq i\leq j\leq n).
 \]
 Define an $n\times n$ symmetric `almost tridiagonal' matrix $B=(B_{ij})$
 by
 \[
    B_{ij}  =
  B_{ji}  =
  \left\{
   \begin{array}{ll}
    \displaystyle
     -\frac{\vev{\v{b},\v{v}_{i-1},\v{v}_{i+1}}}
     {\vev{\v{b},\v{v}_{i-1},\v{v}_{i}}\vev{\v{b},\v{v}_{i},\v{v}_{i+1}}}, 
     \quad
     & \mbox{if $j=i$},
     \\[2ex]
     \displaystyle
     \frac{1}
     {\vev{\v{b},\v{v}_{i},\v{v}_{i+1}}},
     & \mbox{if $j=i+1$},
     \\[2ex]
     \displaystyle
     \frac{1}
     {\vev{\v{b},\v{v}_{1},\v{v}_{n}}},
     & \mbox{if $i=1$ and $j=n$},
     \\[2ex]
     0, & \mbox{otherwise},
   \end{array}
\right.
 \]
 for $1\leq i\leq j\leq n$. Here $\v{v}_{0}:=\v{v}_{n}$ and
 $\v{v}_{n+1}:=\v{v}_{1}$.
 \begin{itemize}
  \item [(i)] For $1\leq i\leq n$,
	\begin{equation}
	 \label{eq:BvSum}
	  B_{i,i-1}\v{v}_{i-1}+
	  B_{i,i}\v{v}_{i}+
	  B_{i,i+1}\v{v}_{i+1}
	  = \frac{\vev{\v{v}_{i-1},\v{v}_{i},\v{v}_{i+1}}}
	  {\vev{\v{b},\v{v}_{i-1},\v{v}_{i}}\vev{\v{b},\v{v}_{i},\v{v}_{i+1}}}
	  \v{b}.
	\end{equation}
	Here we assume $B_{1,0}:=-B_{1,n}$ and $B_{n,n+1}:=-B_{n,1}$.
	
  \item [(ii)] The matrices $A$ and $B$ are inverse to each other.
 \end{itemize}
\end{proposition}

\Proof (i)~
 Suppose for a moment that $ \v{v}_{i-1},\v{v}_{i},\v{v}_{i+1}$ are
 linearly independent. Expanding $\v{b}$ as $\v{b}=\alpha
 \v{v}_{i-1}+\beta \v{v}_{i}+\gamma \v{v}_{i+1}$, it is easy
 to check that the both sides of (\ref{eq:BvSum}) are equal to
 \[
    \frac{1}
   {\alpha \gamma \vev{\v{v}_{i-1},\v{v}_{i},\v{v}_{i+1}}}
   (\alpha\v{v}_{i-1}+\beta \v{v}_{i}+\gamma \v{v}_{i+1}).
 \]
 Since (\ref{eq:BvSum}) is an equality of a rational functions of
 $\v{v}_{i}$'s and $\v{b}$, (\ref{eq:BvSum}) is true in general by the
 continuity argument.

 \smallskip
 (ii)~ It suffices to verify $BA=I$, entry by entry. For
 $1\leq i<j\leq n$,
 \begin{eqnarray*}
   (BA)_{i,j}
   & = &
   B_{i,i{-}1}A_{i{-}1,j}+B_{i,i}A_{i,j}
   +B_{i,i{+}1} A_{i{+}1,j}
   \\
   &=& \frac{1}{2} B_{i,i{-}1} 
   \vev{\v{b},\v{v}_{i-1},\v{v}_{j}}
   +
   \frac{1}{2} B_{i,i}
   \vev{\v{b},\v{v}_{i},\v{v}_{j}}
   +
   \frac{1}{2} B_{i,i{+}1} 
   \vev{\v{b},\v{v}_{i+1},\v{v}_{j}}
   \\
   &  = &
    \frac{1}{2}\vev{\v{b},(B_{i,i-1}\v{v}_{i-1}+
    B_{i,i}\v{v}_{i}+
    B_{i,i+1}\v{v}_{i+1}),\v{v}_{j}},
 \end{eqnarray*}
 which vanishes by (\ref{eq:BvSum}).  Similarly $(BA)_{ij}=0$ for $1\leq
 j<i\leq n$.  On the other hand, for $1< i< n$,
 \begin{eqnarray*}
  (BA)_{i,i}
  &=& B_{i,i{-}1}A_{i{-}1,i}  +B_{i,i{+}1} A_{i{+}1,i}
  \\ &
  =& B_{i{-}1,i}A_{i{-}1,i}
  +B_{i,i{+}1} A_{i,i{+}1}
  \\
  & 
   = & 
  \frac{1}
     {\vev{\v{b},\v{v}_{i-1},\v{v}_{i}}}
  \cdot \frac{1}{2}
  \vev{\v{b},\v{v}_{i-1},\v{v}_{i}}
   +
  \frac{1}
     {\vev{\v{b},\v{v}_{i},\v{v}_{i+1}}}
  \cdot \frac{1}{2}
  \vev{\v{b},\v{v}_{i+1},\v{v}_{i}}
  \\
  &=&\frac{1}{2}+\frac{1}{2}=1.
 \end{eqnarray*}
 With extra care for signs, the cases of $i=1,n$ are similarly verified. 
\qed

%\bibliographystyle{JHEP}
%\bibliography{zonotope}

\def\cprime{$'$}
\providecommand{\href}[2]{#2}\begingroup\raggedright\endgroup

\end{document}